\documentclass[aip,10pt,nofootinbib]{revtex4-1}
\usepackage{graphicx}
\usepackage{amsmath}
\usepackage{amsfonts}
\usepackage{amssymb}
\usepackage{soul}
\usepackage{dsfont}
\usepackage{hyperref}
\usepackage{amstext}

\date{\today}

\begin{document}

\title{Time evolution of coupled multimode and multiresonator optomechanical systems}

\author{David Edward Bruschi}
\affiliation{Institute for Quantum Optics and Quantum Information - IQOQI Vienna, Austrian Academy of Sciences, Boltzmanngasse 3, 1090 Vienna, Austria}
\affiliation{Faculty of Physics, University of Vienna, Boltzmanngasse 5, 1090 Vienna, Austria}
\email{david.edward.bruschi@gmail.com}

\begin{abstract}
We study the time evolution of bosonic systems where multiple driven bosonic modes of light interact with multiple mechanical resonators through arbitrary, time-dependent, optomechanical-like interactions. We find the analytical expression for the full time evolution of the system and compute the expectation value of relevant quantities of interest. Among the most interesting ones, we are able to compute the first-order quantum bipartite coherence between pairs of subsystems, and the analytical expression for the mixedness induced by the nonlinear interaction in the reduced state of the mechanical oscillators. Finally, we also compare our results with a linearised version of the system, and we find a regime where there are qualitative and quantitative differences in the behavior of some measurable quantities. Our results can therefore be used to describe the full time-evolution of the system, to characterise its nonlinear character and explore the validity of the linearisation approach.
\end{abstract}
%\keywords{Optomechanics}
%\submitto{\njp}
\maketitle

%----------------------------------------------------------------------------------------------------------------------------------------------------------------------------------------------------------------------------%
\section{Introduction}\label{intro}
%----------------------------------------------------------------------------------------------------------------------------------------------------------------------------------------------------------------------------%
Physical systems with many constituents are typically difficult to investigate in depth. Arguably, the most imposing challenge faced when studying such systems is the ability to obtain analytical insight into their dynamics, starting from fundamental equations of the theory that describes them. Classical and quantum many-body physics, statistical mechanics and thermodynamics have been developed to successfully tackle systems with large numbers of constituents. Approaches developed using tools and concepts from these areas of research rarely provide full analytical descriptions of the dynamics of complex systems, but are able to provide important coarse-grained information of key aspects of these systems. Regardless of the success of such approaches, a complete, analytical understanding of any system is highly desirable in order to explain existing features, and predict new ones.

In this work we study a quantum system composed of an arbitrary number of bosons, which we conveniently separate into field modes and mechanical oscillators. We assume that these two sets of bosonic modes interact through a nonlinear Hamiltonian, which can be used to model different physical implementations, such as Fabry--P\'erot cavities with a moving-end mirror \cite{Favero:Karrai:2009}, levitated nano-diamonds \cite{Barker:Shneider:2010,Zhang-qi:Tongcang:2013}, membrane-in-the-middle configurations \cite{Jayich:Sankey:2008} and optomechanical crystals \cite{Eichenfield:Camacho:2009,Safavi-Naeini:Hill:2014}.

We employ techniques developed to decouple the time-evolution operator analytically \cite{Puri:2001,Bruschi:Lee:2013,Bruschi:Xuereb:2018},\footnote{An alternative attempt to study analytically the time evolution of such systems was developed in parallel \cite{Brown:Martinez:2013}.} and we provide a full analytical solution to the time evolution of the system. Our results apply to an arbitrary number of interacting systems with arbitrary time-dependent couplings, and are free from approximations. We employ the decoupling of the time-evolution operator to compute analytically the expectation value of physically interesting quantities, such as the average number of field- and resonator-excitations, and we compute the coherence that is induced between different bosonic modes. Furthermore, we are able to provide an analytical expression for the mixedness of the reduced state of the resonators, which can contribute to the understanding of the nonlinear nature of the dynamics. We also comment on the possibility of adding optical drive (i.e., for laser cooling purposes) to the main Hamiltonian, which leads to a system that cannot be treated with the techniques developed in this work

Finally, we apply our results to a set of simple setup, where one cavity mode interacts with an arbitrary number of operators through a standard optomechanical Hamiltonian. Our analytical results are able to provide us with an insight of the quantum behavior of the resonators in this context, and on the coherence introduced between the different subsystems. In particular, we find a simple expression for the mixedness of the resonators which is always nonzero if the nonlinear coupling is present. We are also able to compare our results with those obtained through a ``linearised'' version of the system, an approach typically taken in most approaches to optomechanics \cite{Aspelmeyer:Kippenberg:2014}. We find qualitative and quantitative differences, which we are able to quantify. We conclude discussing the validity and scope of the techniques.

This work is organised as follows. In \autoref{tools} we introduce the necessary tools and the core Hamiltonian. In \autoref{sub:algebra:techniques} we decouple the time-evolution operator analytically. In \autoref{section:time:evolution:quantities:interest} we compute the time evolution of quantities of interest. In \autoref{section:choice:initial:state} and \autoref{section:application:to:optomechanics} we specialise to a physically-relevant initial state and apply our techniques to a few examples.

%----------------------------------------------------------------------------------------------------------------------------------------------------------------------------------------------------------------------------%
\section{Tools}\label{tools}
%----------------------------------------------------------------------------------------------------------------------------------------------------------------------------------------------------------------------------%
We start this work by introducing the necessary tools. The tools that we employ and develop here are not tied to a particular physical system. The only assumption is that the Hamiltonian of the system corresponds to the Hamiltonian presented below, which can be used to model, for example, a cavity optomechanical system \cite{Aspelmeyer:Kippenberg:2014}. Therefore, we emphasize that the approach and results of this work do not depend on the specific implementation chosen. For this reason, we choose to refer to the system as a cavity optomechanical system for simplicity of presentations, and without loss of generality.

Cavity optomechanics studies the interaction of light confined and matter \cite{Aspelmeyer:Kippenberg:2014}. A typical implementation is that of a cavity with a semitransparent mirror-wall, from which an electromagnetic beam can enter. The beam is in resonance with at least one cavity mode, therefore being ``trapped''. On the other end, a mirror-membrane acts as the second wall of the cavity, fully reflecting light. The membrane can vibrate, which affects the fundamental frequency of the mode. The action of the membrane is then modelled effectively as the coupling of the position of a harmonic oscillator to one or more cavity modes. More membranes can be included, such as semitransparent membranes placed at the antinodes of the stationary cavity modes. When one such membrane is present, the system is known to be in a ``membrane-in-the-middle'' configuration \cite{Jayich:Sankey:2008,Aspelmeyer:Kippenberg:2014}. 

%----------------------------------------------------------------------------------------------------------------------------------------------------------------------------------------------------------------------------%
\subsection{Optomechanical Hamiltonians}\label{optomechanics:section}
%----------------------------------------------------------------------------------------------------------------------------------------------------------------------------------------------------------------------------%
In this work, we consider a bosonic system (which we refer to as an optomechanical system) with an arbitrary number of cavity field modes $\{\hat{a}_n,\hat{a}^\dag_n\}$ and an arbitrary number of mechanical modes $\{\hat{b}_p,\hat{b}^\dag_p\}$, where the creation and annihilation operators satisfy the canonical commutation relations $[\hat{a}_n,\hat{a}^\dag_{n'}]=\delta_{nn'}$ and $[\hat{b}_p,\hat{b}^\dag_{p'}]=\delta_{pp'}$, and all other vanish. We then assume that these modes interact through the generalised nonlinear optomechanical Hamiltonian
\begin{align}\label{main:big:time:dependent:Hamiltonian:to:decouple}
\hat {H}_{\textrm{full}}=&\hat {H}_0 +\sum_p\left[\hbar\,\lambda^{(+)}_p\,\hat{B}^{(+)}_p+ \hbar\,\lambda^{(-)}_p\,\hat{B}^{(-)}_p\right]+\sum_{n,p}\hbar\,g_{np}^{(+)} \hat a^\dagger_n\hat a_n \hat{B}^{(+)}_p+\sum_{n,p}\hbar\,g_{np}^{(-)} \hat a^\dagger_n\hat a_n \,\hat{B}^{(-)}_p,
\end{align}
where $\hat {H}_0:=\sum_n \hbar\,\omega_{\textrm{c},n} \hat a_n^\dagger \hat a_n + \sum_p\hbar\,\omega_{\textrm{m},p} \,\hat b_p^\dagger \hat b_p$ is the free Hamiltonian.

\begin{figure}[ht!]{\includegraphics[width=\textwidth]{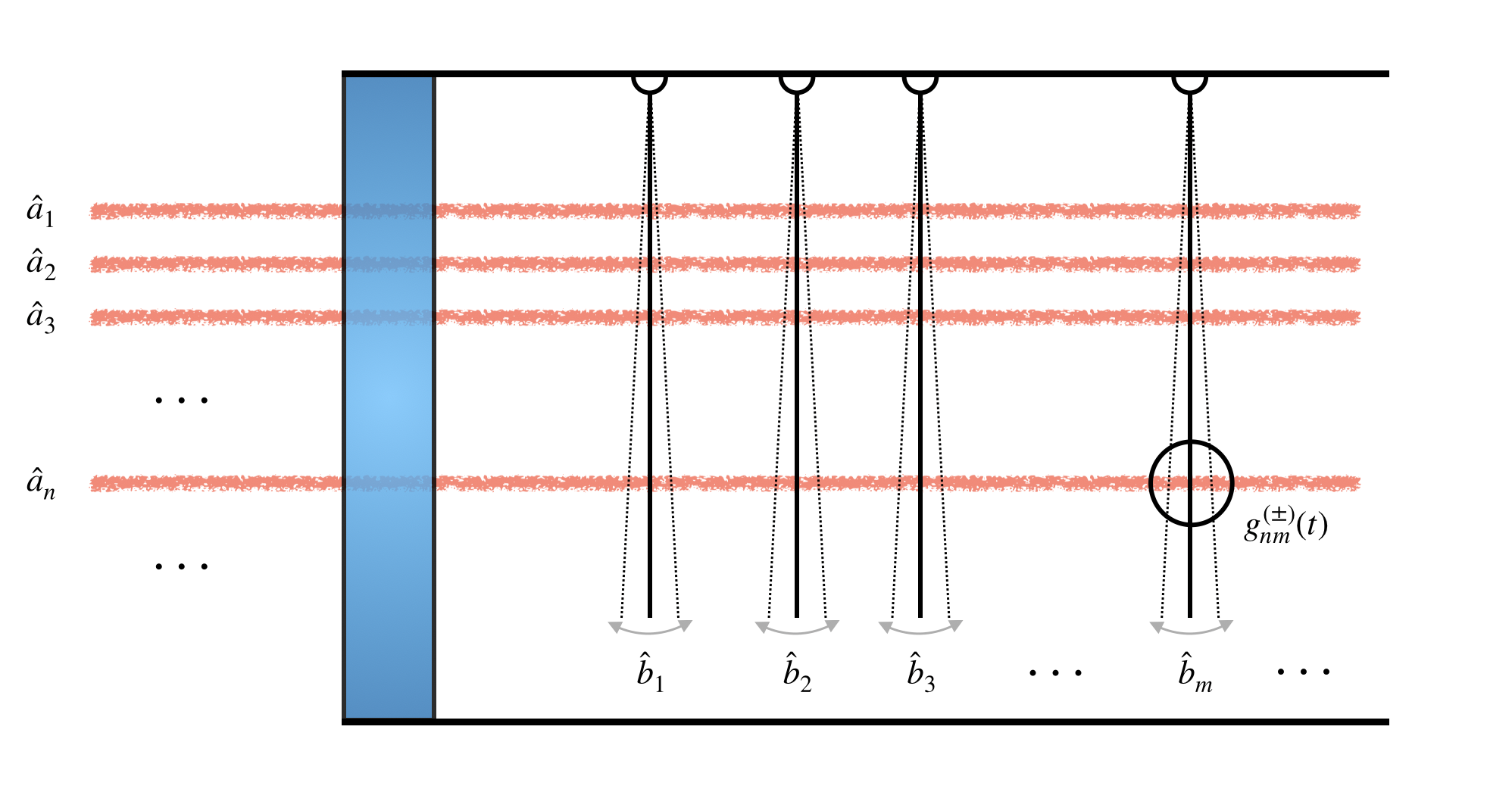}}
\caption{A depiction of the general features of our system. The optical field modes are described by annihilation and creation operators $\hat{a}_n$ and $\hat{a}_n^\dagger$, while the mechanics are described by annihilation and creation operators $\hat{b}_m$ and $\hat{b}_m^\dagger$. The interaction between the cavity mode $n$ and the mechanical mode $m$ is driven by the couplings $g_{nm}^{(\pm)}(t)$. The whole system evolves under the generalised optomechanical-like Hamiltonian \eqref{main:big:time:dependent:Hamiltonian:to:decouple}. Note that there can be an arbitrary number of optical and mechanical degrees of freedom. Finally, although the depiction might recall a cavity-optomechanical implementation, this figure is purely illustrative, since the techniques developed here can be applied to \textit{any} physical implementation that can be described by the Hamiltonian \eqref{main:big:time:dependent:Hamiltonian:to:decouple}.}\label{the:cavity:figure}
\end{figure}

Here we have the cavity mode frequencies $\omega_{\textrm{c},n}$, the mechanical resonator frequencies $\omega_{\textrm{m},p}$, the time dependent couplings $\lambda^{(\pm)}_p(t)$ and $g_{np}^{(\pm)}(t)$, and the Hermitian operators
\begin{align}\label{quadrature:different}
\hat{B}^{(+)}_p=&\hat b_p{}^\dagger+ \hat b_p, & \hat{B}^{(-)}_p=&i\,\left[\hat b_p{}^\dagger- \hat b_p\right].
\end{align} 
The operators defined in \eqref{quadrature:different} can be cast in a more conventional form by noting that they are proportional to the quadrature operators $\hat{x}_{\mathrm{m},p},\hat{p}_{\mathrm{m},p}$ of the mechanical resonators, i.e., $\hat{B}^{(+)}_p\propto \hat{x}_{\mathrm{m},p}$ and $\hat{B}^{(-)}_p\propto \hat{p}_{\mathrm{m},p}$. We retain our choice because the decomposition in terms of creation and annihilation operators is natural to this work. Figure \eqref{the:cavity:figure} illustrates the \textit{general} scheme, that is, it is not a representation of any particular implementation but pictorial description of the system only \eqref{main:big:time:dependent:Hamiltonian:to:decouple} only.

%----------------------------------------------------------------------------------------------------------------------------------------------------------------------------------------------------------------------------%
\subsection{Tackling time evolution of bosonic systems}\label{tools:time:evolution}
%----------------------------------------------------------------------------------------------------------------------------------------------------------------------------------------------------------------------------%
Given a set of $N$ bosonic modes and an arbitrary time-dependent Hamiltonian $\hat{H}(t)$, the unitary time-evolution operator reads 
\begin{align}\label{general:time:evolution:operator}
\hat{U}(t)=\overset{\leftarrow}{\mathcal{T}}\,\exp\left[-\frac{i}{\hbar}\int_0^{t} dt'\,\hat{H}(t')\right],
\end{align}
where $\overset{\leftarrow}{\mathcal{T}}$ is the time ordering operator \cite{Bruschi:Lee:2013}. This expression simplifies dramatically when the Hamiltonian $\hat{H}$ is time independent, in which case one simply has $\hat{U}(t)=\exp[-\frac{i}{\hbar}\,\hat{H}\,t]$. However, we are here interested in general time evolution induced by the time-dependent Hamiltonian \eqref{main:big:time:dependent:Hamiltonian:to:decouple}.  

It is the \textit{main interest of this work} to seek an expression for \eqref{general:time:evolution:operator} of the form
\begin{align}\label{general:time:evolution:operator:decoupled:general}
\hat{U}(t)=\prod_n\,\exp\left[-i\,F_n(t)\,\hat{G}_n\right],
\end{align}
where the $\hat{G}_n$ are time-independent Hermitian operators and the $F_n(t)$ are real time-dependent functions. If the time evolution operator \eqref{general:time:evolution:operator} can be cast in the form \eqref{general:time:evolution:operator:decoupled:general} we say that it has been decoupled. The solution to the formal expression \eqref{general:time:evolution:operator} in terms of its decoupled form \eqref{general:time:evolution:operator:decoupled:general} can be found by employing specifically developed techniques \cite{Bruschi:Lee:2013}, which we outline in \ref{Hamitlonian:decoupling:appendix}. The results are presented in the next section.

%----------------------------------------------------------------------------------------------------------------------------------------------------------------------------------------------------------------------------%
\section{Decoupling the time evolution of the system}\label{sub:algebra:techniques}
%----------------------------------------------------------------------------------------------------------------------------------------------------------------------------------------------------------------------------%
The aim of this work is to show that a decoupling \eqref{general:time:evolution:operator:decoupled:general} exists for the time evolution operator induced by the Hamiltonian \eqref{main:big:time:dependent:Hamiltonian:to:decouple}, and to find an explicit expression for the functions $F_n(t)$. This can be then used to compute the time evolution of any quantity of interest, and of the state of the system.
To achieve our goal, we will employ the techniques developed in \cite{Puri:2001,Bruschi:Lee:2013} to decouple the time-evolution operator $\hat U$ induced the Hamiltonian $\hat {H}$ in \eqref{main:big:time:dependent:Hamiltonian:to:decouple}. More details can be found in \ref{Hamitlonian:decoupling:appendix}. 

We start our task by noting that we can rearrange conveniently the contributing terms to $\hat H(t)$ as $\hat H(t):=\sum_n \hbar\,\omega_{\textrm{c},n} \hat a_n^\dagger \hat a_n+\sum_p\,\hat{H}_p$, where we have introduced the multimode Hamiltonians $\hat{H}_p$ for \emph{fixed} resonator $p$ as
{\small
\begin{align}\label{main:fixed:mode:Hmailtonian}
	\hat {H}_p=& \hbar\,\omega_{\textrm{m},p} \,\hat b_p^\dagger \hat b_p+\hbar\,\lambda^{(+)}_p\,\hat{B}^{(+)}_p+ \hbar\,\lambda^{(-)}_p\,\hat{B}^{(-)}_p+\sum_{n}\hbar\,g_{np}^{(+)} \hat a^\dagger_n\hat a_n \hat{B}^{(+)}_p+\sum_{n}\hbar\,g_{np}^{(-)} \hat a^\dagger_n\hat a_n \,\hat{B}^{(-)}_p.
\end{align}
}
An important property of such Hamiltonians $\hat {H}_p$ is that they commute with each other at all times, i.e., $[\hat {H}_p(t),\hat {H}_{p'}(t')]=0$ for all $p,p'$ and all $t,t'$. This property is important for the next step.

%----------------------------------------------------------------------------------------------------------------------------------------------------------------------------------------------------------------------------%
\subsection{Choice of core Hamiltonian}
%----------------------------------------------------------------------------------------------------------------------------------------------------------------------------------------------------------------------------%
The first step is to employ the fact that $[\hat {H}_p,\hat {H}_{p'}]=0$ for all $p,p'$. This stimulates us to look at an individual Hamiltonian $\hat {H}_p$ and we therefore drop the label $p$ \textit{for this first part of the work}. This leaves us the Hamiltonian
{\small
\begin{align}\label{main:time:dependent:Hamiltonian:to:decouple}
	\hat {H}:=& \hbar\,\omega_{\textrm{m}} \,\hat{N}_b+\hbar\,\lambda^{(+)}\,\hat{B}^{(+)}+ \hbar\,\lambda^{(-)}\,\hat{B}^{(-)}+\sum_{n}\hbar\,g_{n}^{(+)} \hat{N}_n \hat{B}^{(+)}+\sum_{n}\hbar\,g_{n}^{(-)} \hat{N}_n \,\hat{B}^{(-)}.
\end{align}
}
where we have re-defined the operators $\hat{N}_n := \hat a^\dagger_n \hat a_n$, $\hat{N}_b := \hat b^\dagger \hat b$, $\hat{B}^{(+)}:=\hat b{}^\dagger+ \hat b$ and $\hat{B}^{(-)}:=i\,[\hat b{}^\dagger- \hat b]$ for notational convenience.

Decoupling of the Hamiltonian \eqref{main:time:dependent:Hamiltonian:to:decouple} will allow us to decouple the full Hamiltonian \eqref{main:big:time:dependent:Hamiltonian:to:decouple}, as we will show later.
The Hamiltonian \eqref{main:time:dependent:Hamiltonian:to:decouple} is an extension of the standard optomechanical Hamiltonian which, in its simplest form, is obtained from \eqref{main:time:dependent:Hamiltonian:to:decouple} by setting $g_n^{(-)}=\lambda^{(\pm)}=0$, considering only one cavity mode and ignoring the usual external field drive \cite{Aspelmeyer:Kippenberg:2014}. 

%----------------------------------------------------------------------------------------------------------------------------------------------------------------------------------------------------------------------------%
\subsection{Decoupled form of the time-evolution operator}
%----------------------------------------------------------------------------------------------------------------------------------------------------------------------------------------------------------------------------%
We argued that an important achievement that would allow us to understand and control the time evolution of our quantum system is to be able to obtain a decoupled form \eqref{general:time:evolution:operator:decoupled:general} of the time-evolution operator induced by the Hamiltonian \eqref{main:time:dependent:Hamiltonian:to:decouple}. We are able to show that such decoupling exists and has the formal solution
\begin{align}\label{formal:solution:to:the:decoupling}
\hat{U}(t)=&e^{-i\,\sum_n F_n\,\hat{N}_n}\,e^{-i\,F_b\,\hat{N}_b}\,e^{-\frac{i}{2}\,\sum_{nm}\,F_{nm}\,\hat{N}_{nm}}\,e^{-i\,F_+\,\hat{B}^{(+)}}\,e^{-i\,\sum_n\,F_n^{(+)}\,\hat{N}_n\,\hat{B}^{(+)}}\,e^{-i\,F_-\,\hat{B}^{(-)}}\,e^{-i\,\sum_n\,F_n^{(-)}\,\hat{N}_n\,\hat{B}^{(-)}},
\end{align}
and we have introduced the Hermitian operator $\hat{N}_{nm} := \hat a^\dagger_n \hat a_n \hat a^\dagger_m \hat a_m$ for convenience.

The  \emph{time dependent} and \emph{real} functions $F$ can be found explicitly \cite{Bruschi:Lee:2013,Bruschi:Xuereb:2018}, and the full calculations necessary to achieve them are located in \ref{Hamitlonian:decoupling:appendix}.
We find
\begin{align}\label{solution:to:major:differential:equations}
F_b=&\omega_\mathrm{m}\,t\nonumber\\
F_n=&\omega_{\textrm{c},n}\,t-2\,\int_0^t dt'\,\left[\lambda^{(+)}\,\sin(\omega_{\mathrm{m}} t')+\lambda^{(-)}\,\cos(\omega_{\mathrm{m}} t')\right]\,\int_0^{t'} dt''\,\left[g_n^{(+)}\,\cos(\omega_{\mathrm{m}} t'')-g_n^{(-)}\,\sin(\omega_{\mathrm{m}} t'')\right]\nonumber\\
&-2\,\int_0^t dt'\,\left[g_n^{(+)}\,\sin(\omega_{\mathrm{m}} t)+g_n^{(-)}\,\cos(\omega_{\mathrm{m}} t')\right]\int_0^{t'} dt''\,\left[\lambda^{(+)}\,\cos(\omega_{\mathrm{m}} t'')-\lambda^{(-)}\,\textrm{s}(t'')\right]\nonumber\\
F_{nm}=&-4\,\int_0^t dt'\,\left[g_m^{(+)}\,\sin(\omega_{\mathrm{m}} t')+g_m^{(-)}\,\cos(\omega_{\mathrm{m}} t')\right]\int_0^{t'} dt''\,\left[g_n^{(+)}\,\cos(\omega_{\mathrm{m}} t'')-g_n^{(-)}\,\sin(\omega_{\mathrm{m}} t'')\right]\nonumber\\
F_+ =& \int_0^t dt'\,\left[\lambda^{(+)}\,\cos(\omega_{\mathrm{m}} t')-\lambda^{(-)}\,\sin(\omega_{\mathrm{m}} t)\right]
\nonumber\\
F_- =& -\int_0^t dt'\,\left[\lambda^{(+)}\,\sin(\omega_{\mathrm{m}} t')+\lambda^{(-)}\,\cos(\omega_{\mathrm{m}} t')\right]
\nonumber\\
F_n^{(+)} =& \int_0^t dt'\,\left[g_n^{(+)}\,\cos(\omega_{\mathrm{m}} t')-g_n^{(-)}\,\sin(\omega_{\mathrm{m}} t')\right]
\nonumber\\
F_n^{(-)} =& -\int_0^t dt'\,\left[g_n^{(+)}\,\sin(\omega_{\mathrm{m}} t')+g_n^{(-)}\,\cos(\omega_{\mathrm{m}} t')\right].
\end{align}
We would like to emphasise that the existence of an analytical expression for the $F$ functions is remarkable in its own right. Notwithstanding the fact that the set of necessary operators \eqref{basis:operator:Lie:algebra} is infinite, and therefore the number of terms to be expected in the decoupled form \eqref{general:time:evolution:operator:decoupled:general} is also infinite, it is possible to obtain an explicit expression of the time evolution through \eqref{solution:to:major:differential:equations} given an explicit expression for the time dependent couplings $\lambda_\pm$ and $g_n^{(\pm)}$.

This achievement should also shed light on the reasons why it has been difficult, so far, to fully characterise how such a nonlinear system evolves in time. A naive application of the Baker-Campbell-Hausdorff formula would have not been able to provide us  in an efficient way with the full set \eqref{basis:operator:Lie:algebra} of necessary operators, which in turn allowed us to compute the functions \eqref{solution:to:major:differential:equations}. It should therefore be clear now why the techniques used and developed here can lead to a deeper understanding of the dynamics of nonlinear quantum systems.

%----------------------------------------------------------------------------------------------------------------------------------------------------------------------------------------------------------------------------%
\subsection{Full multiresonator nonlinear decoupled solution}
%----------------------------------------------------------------------------------------------------------------------------------------------------------------------------------------------------------------------------%
We have decoupled analytically the time-evolution operator induced by the Hamiltonian \eqref{main:time:dependent:Hamiltonian:to:decouple}. This has led us to the analytical expressions  \eqref{solution:to:major:differential:equations}. We recall that there is a Hamiltonian 
\eqref{main:time:dependent:Hamiltonian:to:decouple} for each mechanical mode $\hat{b}_p$. Therefore, we can reconstruct the \emph{decoupled} time-evolution operator $\hat{U}_{\textrm{full}}$ induced by the Hamiltonian $\hat{H}_{\textrm{full}}$ in \eqref{main:big:time:dependent:Hamiltonian:to:decouple} by recalling that $[\hat{H}_p,\hat{H}_{p'}]=0$ for all $p,p'$. Some algebra leads us to
\begin{align}\label{solution:to:the:decoupling:of:full:nonlinear:part}
\hat{U}_{\textrm{full}}(t)=&e^{-i\,\left[\sum_n \omega_{\textrm{c},n}\,t+\sum_{np} \tilde{F}_{\textrm{c},n}^{(p)}\right]\hat{N}_n}\,e^{-i\,\sum_p F_{\textrm{m}}^{(p)}\hat{b}^\dag_p \hat{b}_p}\,e^{-\frac{i}{2} \sum_{nmp} F_{nm}^{(p)} \hat{N}_{nm}}\,e^{-i\,\sum_p F_+^{(p)}\hat{B}^{(p,+)}}\,e^{-i\,\sum_{np} F_n^{(p,+)}\hat{N}_n \hat{B}^{(p,+)}}\nonumber\\
&\times e^{-i\,\sum_p F_-^{(p)} \hat{B}^{(p,-)}}\,e^{-i\,\sum_{np} F_n^{(p,-)} \hat{N}_n \hat{B}^{(p,-)}},
\end{align}
where the updated functions read
\begin{align}\label{solution:to:major:differential:equations:with:all:resonators}
F_{\textrm{m}}^{(p)}&=\omega_{\mathrm{m},p}\,t\nonumber\\
\tilde{F}_{\textrm{c},n}^{(p)}&=-2\,\int_0^t dt'\,\left[\lambda_p^{(+)}\,\sin(\omega_{\mathrm{m},p} t')+\lambda_p^{(-)}\,\cos(\omega_{\mathrm{m},p} t')\right]\,\int_0^{t'} dt''\,\left[g_{np}^{(+)}\,\cos(\omega_{\mathrm{m},p} t'')-g_{np}^{(-)}\,\sin(\omega_{\mathrm{m},p} t'')\right]\nonumber\\
&-2\,\int_0^t dt'\,\left[g_{np}^{(+)}\,\sin(\omega_{\mathrm{m},p} t')+g_{np}^{(-)}\,\cos(\omega_{\mathrm{m},p} t')\right]\int_0^{t'} dt''\,\left[\lambda^{(+)}\,\cos(\omega_{\mathrm{m},p} t'')-\lambda^{(-)}\,\sin(\omega_{\mathrm{m},p} t'')\right]\nonumber\\
F_{nm}^{(p)}&=-4\,\int_0^t dt'\,\left[g_{mp}^{(+)}\,\sin(\omega_{\mathrm{m},p} t')+g_{mp}^{(-)}\,\cos(\omega_{\mathrm{m},p} t')\right]\int_0^{t'} dt''\,\left[g_{np}^{(+)}\,\cos(\omega_{\mathrm{m},p} t'')-g_{np}^{(-)}\,\sin(\omega_{\mathrm{m},p} t'')\right]\nonumber\\
F_+^{(p)} &= \int_0^t dt'\,\left[\lambda^{(+)}\,\cos(\omega_{\mathrm{m},p} t')-\lambda^{(-)}\,\sin(\omega_{\mathrm{m},p} t')\right]
\nonumber\\
F_-^{(p)} &= -\int_0^t dt'\,\left[\lambda^{(+)}\,\sin(\omega_{\mathrm{m},p} t')+\lambda^{(-)}\,\cos(\omega_{\mathrm{m},p} t')\right]
\nonumber\\
F_n^{(p,+)} &= \int_0^t dt'\,\left[g_{np}^{(+)}\,\cos(\omega_{\mathrm{m},p} t')-g_{np}^{(-)}\,\sin(\omega_{\mathrm{m},p} t')\right]
\nonumber\\
F_n^{(p,-)} &= -\int_0^t dt'\,\left[g_{np}^{(+)}\,\sin(\omega_{\mathrm{m},p} t')+g_{np}^{(-)}\,\cos(\omega_{\mathrm{m},p} t')\right].
\end{align}
As noted before, we find it remarkable that the time evolution for this system can be decoupled exactly and analytically. Our ability to find such solution is a consequence of the structure of the full operator algebra, which allows for the use of many ``tricks'' to our advantage~\cite{Puri:2001,Bruschi:Lee:2013}. Therefore, we believe that this approach has shown a clear and marked advantage with respect to a more brute-force application of the Baker-Campbell-Hausdorff formula.

%----------------------------------------------------------------------------------------------------------------------------------------------------------------------------------------------------------------------------%
\subsection{Time evolution of optically driven multimode and multiresonator systems}
%----------------------------------------------------------------------------------------------------------------------------------------------------------------------------------------------------------------------------%
We would like to extend the results obtained above to tackle the full time evolution $\hat U(t)$ induced by an optomechanical Hamiltonian $\hat{H}$ that includes an external optical drive (potentially for each mode) in addition the interactions already considered in \eqref{main:big:time:dependent:Hamiltonian:to:decouple}.\footnote{It is a common practice to employ such a drive in physical implementations, for purposes such as cooling.} We therefore would like to consider the Hamiltonian
\begin{align}\label{main:big:time:dependent:problematic:Hamiltonian:to:decouple}
\hat {H}=& \hat{H}_{\textrm{full}}+\sum_n\left[\hbar\,\xi^{(+)}_n\,\hat{A}^{(+)}_n+ \hbar\,\xi^{(-)}_n\,\hat{A}^{(-)}_n\right].
\end{align}
Here we have introduced the operators $\hat{A}^{(+)}_n=\hat a_n{}^\dagger+ \hat a_n$ and $\hat{A}^{(-)}_n=i\,\left[\hat a_n{}^\dagger- \hat a_n\right]$, and the time-dependent couplings $\xi^{(\pm)}_n(t)$. Note that the operators $\hat{A}^{(\pm)}_n$ are proportional to the quadrature operators $\hat{x}_{\mathrm{c},n},\hat{p}_{\mathrm{c},n}$ of the cavity modes, i.e., $\hat{A}^{(+)}_n\propto \hat{x}_{\mathrm{c},n}$, $\hat{A}^{(-)}_n\propto \hat{p}_{\mathrm{c},n}$.

The Hamiltonian \eqref{main:big:time:dependent:problematic:Hamiltonian:to:decouple} has the form $\hat {H}= \hat {H}_{\textrm{full}}+\hat {H}_{\textrm{Dr}}$, where we have decided to define $\hat {H}_{\textrm{Dr}}=\sum_{n}\left[\hbar\,\xi^{(+)}_n\,\hat{A}^{(+)}_n+ \hbar\,\xi^{(-)}_n\,\hat{A}^{(-)}_n\right]$ for convenience.
The time-evolution operator $\hat U(t)$ induced by $\hat {H}$ has the expression \eqref{general:time:evolution:operator}, which can be manipulated to obtain the equivalent expression
\begin{align}\label{general:full:time:evolution:operator:second:expression}
\hat U(t)=&\hat U_{\textrm{full}}(t)\,\overset{\leftarrow}{\mathcal{T}}\,\exp\left[-\frac{i}{\hbar}\int_0^{t} dt'\, U_{\textrm{full}}^\dag(t')\,\hat{H}_{\textrm{Dr}}(t')\, U_{\textrm{full}}(t')\right],
\end{align}
where $\hat U_{\textrm{full}}(t):=\overset{\leftarrow}{\mathcal{T}}\,\exp[-\frac{i}{\hbar}\int_0^{t} dt'\,\hat{H}_{\textrm{full}}(t')]$.
Since $[\hat{H}_p,\hat{H}_{p'}]=0$, we obtain
\begin{align}\label{final:general:to:still:solve}
\hat U(t)=&U_{\textrm{full}}(t)\,\overset{\leftarrow}{\mathcal{T}}\,\exp\left[-i\,\sum_n\int_0^{t} dt'\, \hat U^\dag_{\textrm{full}}(t')\,\left(\xi_n\,\hat{a}^\dag_n+\xi^*_n\,\hat{a}_n\right)\,\hat U_{\textrm{full}}(t')\right],
\end{align}
where we have defined the complex coupling $\xi_n:=\xi_n^{(+)}+i\,\xi_n^{(-)}$.

Unfortunately, we cannot proceed any further with simplifications of \eqref{final:general:to:still:solve}. The explicit expression in the time-ordered exponential leads to a time ordered exponential of exponential operators, which cannot be treated with the tools described here. We leave it to further work to study this important case in more detail.

%----------------------------------------------------------------------------------------------------------------------------------------------------------------------------------------------------------------------------%
\section{Time evolution of quantities of interest}\label{section:time:evolution:quantities:interest}
%----------------------------------------------------------------------------------------------------------------------------------------------------------------------------------------------------------------------------%
The decoupling achieved above allows us to obtain analytical control on the time evolution of the system. This includes our ability to compute explicitly the expectation value of many quantities of interest, which we proceed to do below. 

%----------------------------------------------------------------------------------------------------------------------------------------------------------------------------------------------------------------------------%
\subsection{Time evolution of quantities of interest: mode operators}
%----------------------------------------------------------------------------------------------------------------------------------------------------------------------------------------------------------------------------%
We start by computing the time evolution of the mode operators $\hat{a}_k$ and $\hat{b}_p$, through which one can compute the time evolution of most quantities of interest.
Their time dependence is induced by the nonlinear Hamiltonian $\hat{H}_{\textrm{full}}$ and is obtained through the standard Heisenberg equation as $\hat{a}_k(t):=\hat{U}_{\textrm{full}}^\dag\,\hat{a}_k\,\hat{U}_{\textrm{full}}$ and $\hat{b}_p(t):=\hat{U}_{\textrm{full}}^\dag\,\hat{b}_p\,\hat{U}_{\textrm{full}}$ respectively. After some algebra we find
\begin{align}\label{time:evolution:of:basic:operators}
	\hat{a}_k(t) =& e^{-i\,\omega_{\textrm{c},k}\,t}e^{-i\,\sum_p\,\left(F_{kk}^{(p)}+\tilde{F}_{\textrm{c},k}^{(p)}+2\,F^{(p,+)}_k\,F_-^{(p)}\right)}\,e^{-i\,\sum_{p,n}\,\left(F^{(p)}_{\{kn\}}+2\,F^{(p,+)}_k\,F^{(p,-)}_n\right)\hat{N}_n}\,e^{-i\,\sum_p\,F^{(p,+)}_k\,\hat{B}^{(p,+)}}e^{-i\,\sum_p\,F^{(p,-)}_k\,\hat{B}^{(p,-)}}\hat{a}_k\nonumber\\
	\hat{b}_p(t) =& e^{-i\,F^{(p)}_{\textrm{m}}}\,\left[\hat{b}_p-i\,F^{(p)}-i\,\sum_n\,F^{(p)}_n\,\hat{N}_n\right],
\end{align}
where we have defined $F^{(p)}(t):=F^{(p)}_-(t)+i\,F^{(p)}_+(t)$ and $F^{(p)}_n(t):=F^{(p,+)}_n(t)+i\,F^{(p,-)}_n(t)$, and the notation $F^{(p)}_{\{nm\}}:=\frac{1}{2}[F^{(p)}_{nm}+F^{(p)}_{mn}]$.

%----------------------------------------------------------------------------------------------------------------------------------------------------------------------------------------------------------------------------%
\subsection{Final reduced state of the mechanical resonators}
%----------------------------------------------------------------------------------------------------------------------------------------------------------------------------------------------------------------------------%
We continue by computing the final reduced state $\hat{\rho}_{\textrm{m}}(t)$ of the mechanical resonators. The reduced state is defined by $\hat{\rho}_{\textrm{m}}(t):=\textrm{Tr}_{\textrm{Phot}}(\hat{\rho}_{\textrm{NL}}(t))$, that is, by tracing over all of the cavity modes.
In \ref{reduced:resonator:state:section} we provide all of the detailed computations for this part. We assume that the initial state $\hat{\rho}_{\mathrm{c}}(0)$ of the cavity modes is separable from the initial state $\hat{\rho}_{\mathrm{m}}(0)$ of the mechanical modes. This implies that the full initial state is $\hat{\rho}_0=\hat{\rho}_{\mathrm{c}}(0)\otimes\hat{\rho}_{\mathrm{m}}(0)$. It is not difficult to show that the reduced state $\hat{\rho}_{\textrm{m}}(t)$ at time $t$ has the form
\begin{align}\label{nonlinear:redced:mechanical:state}
\hat{\rho}_{\textrm{m}}(t)=\sum_{\overset{\{n_k\}}{k\in\mathcal{I}}} p_{\{n_k\}}\,\hat{D}_{\{n_k\}}\,\hat{\rho}_{\textrm{m}}(0)\,\hat{D}_{\{n_k\}}^\dag,
\end{align}
where we have introduced $\sum_{\overset{\{n_k\}}{k\in\mathcal{I}}}:=\sum_{n_1,n_2,...,n_N}$ for $N$ modes that belong to the set of all possible combinations of excitations $\mathcal{I}$, while $\sum_{k\in\mathcal{I}}\,J_k:=J_1+J_2+...+J_N$ for any $k$-dependent quantities $J_k$. We have also introduced
\begin{align}\label{d:expression}
p_{\{n_k\}}:=&\langle n_1,...,n_N|\hat{\rho}_{\mathrm{c}}(0)| n_1,...,n_N\rangle\nonumber\\
\hat{D}_{\{n_k\}}:=&e^{-i\sum_p\,F^{(p)}_\textrm{m}\,\hat{b}^\dag_p\hat{b}_p}\,e^{-i\sum_p\left(F^{(p)}_++\sum_{k\in\mathcal{I}}n_k\,F^{(p,+)}_k\right)\,\hat{B}^{(p,+)}}\,e^{-i\sum_p\left(F^{(p)}_-+\sum_{k\in\mathcal{I}}n_k\,F^{(p,-)}_k\right)\,\hat{B}^{(p,-)}}.
\end{align}
Note that we have $\textrm{Tr}(\hat{\rho}_{\textrm{m}}(t))=\sum_{\overset{\{n_k\}}{k\in\mathcal{I}}} p_{\{n_k\}}=1$ as expected.

%----------------------------------------------------------------------------------------------------------------------------------------------------------------------------------------------------------------------------%
\subsection{Mode population}
%----------------------------------------------------------------------------------------------------------------------------------------------------------------------------------------------------------------------------%
The time evolution of the modes allows us to immediately compute the operators that ``count'' the number of excitations at any moment in time, namely $\hat{a}_k^\dag(t)\hat{a}_k(t)$ and $\hat{b}_k^\dag(t)\hat{b}_k(t)$. They read
\begin{align}\label{time:evolution:of:number:operators}
	\hat{a}_k^\dag(t)\hat{a}_k(t) =& \hat{a}_k^\dag\hat{a}_k\nonumber\\
	\hat{b}_k^\dag(t)\hat{b}_k(t) =& \hat{b}_k^\dag\hat{b}_k-i\left[F^{(k)}+\sum_n\,F^{(k)}_n\,\hat{N}_n\right]\, \hat{b}_k^\dag+i\left[F^{(k)*}+\sum_nF^{(k)*}_n\hat{N}_n\right] \hat{b}_k+F^{(k)}\,\sum_n\,F^{(k)*}_n\,\hat{N}_n+F^{(k)*}\,\sum_n\,F^{(k)}_n\,\hat{N}_n\nonumber\\
	&+|F^{(k)}|^2+\sum_{nm}\,F^{(k)}_n\,F^{(k)*}_m\,\hat{N}_n\,\hat{N}_m.
\end{align}
The number operator $\hat{a}_k^\dag(t)\hat{a}_k(t)$ of each field mode is a conserved quantity as can be immediately seen from the Hamiltonian \eqref{main:big:time:dependent:Hamiltonian:to:decouple} (that is, it commutes with the whole Hamiltonian). However, the number operators $\hat{b}_k^\dag(t)\hat{b}_k(t)$ of the resonators are not, and they depend on the nonlinear coupling through the functions $F^{(k)}$ and $F^{(k)}_n$. 

We note here that, if the nonlinear couplings are small, i.e.,  they are proportional to $\epsilon\ll1$, we have that $F^{(k)}\sim\epsilon$, $F^{(k)}_n\sim\epsilon$ and $F^{(k)}_n\,F^{(k)*}_m\sim\epsilon^2$. Therefore, the last term contributing to the number operator $\hat{b}_k^\dag(t)\hat{b}_k(t)$ in \eqref{time:evolution:of:number:operators} is a negligible contribution in this regime. When this occurs, it is easy to check that the result is equivalent to what would be obtained through first-order perturbation theory, as expected.
In this sense, the operators \eqref{time:evolution:of:number:operators} inform us on the full evolution of the resonator's population and contain a signature of the full nonlinear character of the system.

%----------------------------------------------------------------------------------------------------------------------------------------------------------------------------------------------------------------------------%
\subsection{First-order bipartite quantum coherence}
%----------------------------------------------------------------------------------------------------------------------------------------------------------------------------------------------------------------------------%
Given two modes $m$ and $n$, we call the correlation $\langle\hat{d}^\dag_m\hat{d}_n\rangle$ the (first-order) bipartite coherence, sometimes denoted by $G^{(1)}_{mn}$ in optics \cite{Walls:Milburn:2001,Kalaga:Kowalewska:2016}. This definition applies in general to any state. This measure corresponds to a simple interferometric setup, where we collect the photons in the modes $m$ and $n$, add a phase difference between their paths, and let them interfere.
We will witness the formation of an interference pattern only if the quantity $\langle\hat{d}^\dag_m\hat{d}_n\rangle$ is non-zero. This quantity can be normalized by the power in each mode, and in this case we recover the standard definition of first-order amplitude correlation function $g^{(1)}_{mn}$ from quantum optics
applied to modes $m$ and $n$, namely
\begin{align}\label{first-order:amplitude:correlation:function}
	g^{(1)}_{mn}(t):=\frac{|\langle\hat{d}^\dag_m(t)\hat{d}_n(t)\rangle|}{\sqrt{\langle\hat{d}^\dag_m(t)\hat{d}_m(t)\rangle\langle\hat{d}^\dag_n(t)\hat{d}_n(t)\rangle}}.
\end{align}
It is not difficult to employ our results and compute $g^{(1)}_{mn}$ for pairs of cavity modes $\hat{a}_k$ and $\hat{a}_{k'}$, for pairs of resonator modes $\hat{b}_k$ and $\hat{b}_{k'}$, or for pairs of cavity and resonator modes $\hat{a}_k$ and $\hat{b}_{k'}$.
The results can be obtained analytically but are not illuminating and we omit to print the general formulas. Instead, we will give a few explicit results later on, when the initial state has been chosen.

%----------------------------------------------------------------------------------------------------------------------------------------------------------------------------------------------------------------------------%
\subsection{Mixedness and linear entropy}
%----------------------------------------------------------------------------------------------------------------------------------------------------------------------------------------------------------------------------%
The final quantity that we are interested in computing is the amount of mixedness induced in the two main subsystems, i.e., the optical and mechanical one, due to the interaction between these systems. In particular, we would like to be able to isolate the contribution to any such induced mixedness and the nonlinear part of the interaction. 

A measure of the mixedness of a state is the \emph{linear entropy} $S_N$, defined as $S_N=1-\textrm{Tr}(\hat{\rho}^2)$ and vanishes for pure states~\cite{Peters:Wei:2004}.\footnote{Any pure state $\hat{\rho}$ satisfies $\hat{\rho}^2=\hat{\rho}$. Since $\textrm{Tr}(\hat{\rho})=1$ this implies that $\textrm{Tr}(\hat{\rho}^2)=1$. This motivates the definition of the linear entropy. We also note that, contrary to the case of finite-dimensional systems with dimension $d$, where the state $\rho$ with maximal mixedness $S_N=1/d$ is the diagonal state with uniform eigenvalues $\lambda=1/d$, i.e., $\hat{\rho}=\sum_n \lambda |n\rangle\langle n|$  for an complete orthonormal basis $|n\rangle$, the case of infinite dimensional systems is more subtle. Clearly, there cannot exist a diagonal state with uniform eigenvalues $\lambda=1/d$, since d is infinite. However, one could have diagonal states with \emph{finite} amount of uniform, non-zero eigenvalues $\lambda=1/k$, were $k$ can be arbitrary. In this case, one would still have $S_N=1/k$, which can be made arbitrarily small by increasing $k$.} 
The time evolution of the system is expected to induce coherence between the photonic part and the mechanical part and the coherence will induce mixedness between these parts, which we can quantify using the linear entropy $S_N$. We focus on the reduced state $\hat{\rho}_{\textrm{m}}(t)$ of the mechanical subsystem obtain above \eqref{nonlinear:redced:mechanical:state}.
In our case we have 
\begin{align}\label{general:expression:linear:entropy}
S_N=&1-\textrm{Tr}\left(\hat{\rho}_{\textrm{m}}^2(t)\right)\nonumber\\
%=&1-\sum_{\overset{\{n_k,m_{k}\}}{k\in\mathcal{I}}} p_{\{n_k\}}\,p_{\{m_k\}}\,\textrm{Tr}\left(\hat{D}_{\{n_k\}}\,\rho_{\textrm{m}}(0)\,\hat{D}_{\{n_k\}}^\dag\,\hat{D}_{\{m_{k}\}}\,\rho_{\textrm{m}}(0)\,\hat{D}_{\{m_{k}\}}^\dag\right)\nonumber\\
=&1-\sum_{\overset{\{n_k,m_{k}\}}{k\in\mathcal{I}}} p_{\{n_k\}}\,p_{\{m_k\}}\,\textrm{Tr}\left(\hat{D}_{\{m_{k}\}}^\dag\,\hat{D}_{\{n_k\}}\,\rho_{\textrm{m}}(0)\,\hat{D}_{\{n_k\}}^\dag\,\hat{D}_{\{m_{k}\}}\,\rho_{\textrm{m}}(0)\right),
\end{align}
where it is easy to check that $\hat{D}_{\{n_k\}}^\dag\,\hat{D}_{\{m_{k}\}}
=\prod_p e^{i\theta_p'}\,\exp\left[i\,\Delta^{(p)}_{\{n_k,m_{k}\}}\,\hat{b}^\dag_p+\textrm{h.c.}\right]$. Here we have introduced the very useful functions $F^{(p)}_k:=F^{(p,+)}_k+i\,F^{(p,-)}_k$ and $\Delta^{(p)}_{\{n_k,m_{k}\}}:=\sum_{k\in\mathcal{I}}\left(n_k-m_k\right)F^{(p)}_k$. The exact expression of the phase $e^{i\,\theta'}$ is irrelevant since it clearly cancels out in \eqref{general:expression:linear:entropy}. 

We know that when all the nonlinear couplings vanish, i.e., $g^{(\pm)}_{np}=0$, we have $F^{(p)}_k=0$. As an immediate consequence of this is that the expression \eqref{general:expression:linear:entropy} reduces to $S_N(t)=1-\textrm{Tr}(\hat{\rho}^2_{\textrm{m}}(0))=S_N(0)$. Therefore, the mixedness introduced in the reduced state $\hat{\rho}_{\textrm{m}}$ of the resonators is a \textit{direct} and \textit{only} consequence of the nonlinear interaction. This analytical insight is, of course, perfectly in line with what is expected.

%----------------------------------------------------------------------------------------------------------------------------------------------------------------------------------------------------------------------------%
\section{Application to initial coherent state of the cavity modes and thermal state of the mechanical modes}\label{section:choice:initial:state}
%----------------------------------------------------------------------------------------------------------------------------------------------------------------------------------------------------------------------------%
Here we apply our results to a more concrete setup. We assume that there are a limited amount of modes $k\in\mathcal{I}$ that are initially in a coherent state $|\mu_k\rangle$ with parameter $\mu_k$ and defined by $\hat a_k|\mu_k\rangle=\mu_k\,|\mu_k\rangle$, while the cavity modes $s\notin\mathcal{I}$ are each in their respective vacuum state $|0\rangle_s$. 

We also assume that the mechanical modes $\hat{b}_p$ are initially in a thermal state $\hat{\rho}_{\textrm{m}}(0)=\prod_{p\notin\mathcal{I}}\sum_{j_p}\frac{\tanh^{2\,j_p}(r_p)}{\cosh^2(r_p)}|j_p\rangle\langle j_p|$, with temperature $T$ and parameter $r_p$ defined by $\tanh(r_p):=\exp[-\frac{\hbar\,\omega_{\textrm{m},p}}{2\,k_\mathrm{B}\,T}]$. This is the standard initial setup in most applications, such as those with mechanical oscillators \cite{Teufel:Donner:2011} and with levitated nano-objects \cite{Jain:Gieseler:2016}. 
Note that the set $\mathcal{I}$ might include any number $N$ of modes with $N\geq1$, and that $N_{\textrm{i},p}\equiv\sinh^2(r_p)$ is the initial population of thermal mechanical phonons.\footnote{The temperatures can be lowered to values that allow to reduce the number $N_{\textrm{i},p}$ of initial thermal phonons to $N_{\textrm{i},p}\sim0.34$, in the case of mechanical oscillators \cite{Teufel:Donner:2011}. In the case of levitated nano-objects \cite{Jain:Gieseler:2016}, one can reach temperatures that give rise to an average population of $N_{\textrm{i},p}\sim60$.}

The initial state of the system $\hat{\rho}(0)$ is then separable in the mode/resonator bipartition, and has the expression 
\begin{align}\label{initial:state:optomechanics}
	\hat{\rho}(0) =& \prod_{k\in\mathcal{I}}|\mu_k\rangle\langle\mu_k|\otimes \prod_{s\notin\mathcal{I}}|0\rangle\langle0|_s\otimes \hat{\rho}_{\textrm{m}}(0).
\end{align}
Note that the state is mixed due to the initial temperature in the mechanical modes.

%----------------------------------------------------------------------------------------------------------------------------------------------------------------------------------------------------------------------------%
\subsection{Final reduced state of the mechanical resonators}
%----------------------------------------------------------------------------------------------------------------------------------------------------------------------------------------------------------------------------%
The final reduced state $\hat{\rho}_{\mathrm{m}}(t)$ of the mechanical modes has been computed for the general case and reads \eqref{nonlinear:redced:mechanical:state}. In the specific case we are studying here, we have to use $p_{\{n_k\}}=\prod_{k\in\mathcal{I}}\frac{|\mu_k|^{2\,n_k}}{n_k!}\,e^{-|\mu_k|^2}$ and we will obtain the final expression. This expression is cumbersome and we avoid printing it here.

%----------------------------------------------------------------------------------------------------------------------------------------------------------------------------------------------------------------------------%
\subsection{First-order bipartite coherence}
%----------------------------------------------------------------------------------------------------------------------------------------------------------------------------------------------------------------------------%
In this case, we can provide some explicit formulas given that we have specified the initial state of the system. We use the expression \eqref{first-order:amplitude:correlation:function} and we provide expressions for the nominator and denominator separately. 

To compute the first-order coherence between two resonators $\hat{b}_k$ and $\hat{b}_{k'}$ or a cavity mode $\hat{a}_k$ and resonator $\hat{b}_{k'}$ we need the following on-diagonal expressions
\begin{align}\label{expressions:for:the:first:order:bipartite:coherence:on:diagonal}
\langle\hat{a}_k^\dag\,\hat{a}_k\rangle=&|\mu_k|^2\nonumber\\
\langle\hat{b}_{p}^\dag\hat{b}_{p}\rangle=&N_{\textrm{i},p}+ |F^{(p)}|^2-2\,\sum_n\Re(F^{(p)}F^{(p)*}_n)|\mu_n|^2+\sum_n|F^{(p)}_n|^2(|\mu_n|^2+|\mu_n|^4)+2\,\sum_{n> m}\Re(F^{(p)}_nF^{(p)*}_m)\,|\mu_n|^2\,|\mu_m|^2,
\end{align}
and off-diagonal expressions
\begin{align}\label{expressions:for:the:first:order:bipartite:coherence:off:diagonal}
|\langle\hat{a}_{k}^\dag\hat{a}_{k'}\rangle|=&\exp\left[-2\,\sum_m\,|\mu_{m}|^2\,\sin^2\left(\frac{1}{2}\,\sum_p\left(F^{(p)}_{\{km\}}-F^{(p)}_{\{k'm\}}+2\,(F^{(p,+)}_k-F^{(p,+)}_{k'})\,F^{(p,-)}_{m}\right)\right)\right]\nonumber\\
&\times\, |\mu_k|\,|\,\mu_{k'}|\,e^{-\frac{1}{2}\,\sum_p\cosh(2\,r_p) |F_k^{(p)}-F_{k'}^{(p)}|^2}\nonumber\\
|\langle\hat{a}_{k}^\dag\hat{b}_{p'}\rangle|=&\exp\left[-2\,\sum_m\,|\mu_{m}|^2\,\sin^2\left(\frac{1}{2}\,\sum_p\left(F^{(p)}_{\{km\}}+2\,F^{(p,+)}_k\,F^{(p,-)}_{m}\right)\right)\right]\nonumber\\
&\times\,\left|F_{k}^{(p')}\,N_{\textrm{i},p'}-F^{(p')}-\sum_{n} F^{(p')}_n\,e^{i\sum_p\left(F^{(p)}_{\{kn\}}+2\,F^{(p,+)}_k\,F^{(p,-)}_n\right)}\,|\mu_n|^2\right|\nonumber\\
&\times|\mu_k|\,e^{-\frac{1}{2}\,\sum_p\cosh(2\,r_p) |F_k^{(p)}|^2}\nonumber\\
|\langle\hat{b}_{p}^\dag\hat{b}_{p'}\rangle|=& \left|F^{(p)*}\,F^{(p')}+\sum_n F^{(p)}F^{(p')*}_n|\mu_n|^2+\sum_n F^{(p')*}F^{(p)}_n|\mu_n|^2\right.\nonumber\\
&\left.+\sum_nF^{(p)*}_n\,F^{(p')}_n(|\mu_n|^2+|\mu_n|^4)+\sum_{n\neq m}F^{(p)*}_n\,F^{(p')}_m\,|\mu_n|^2\,|\mu_m|^2\right|,
\end{align}
all of which have been computed  using the quantities in \ref{useful:expressions}. In particular, to obtain the expressions \eqref{expressions:for:the:first:order:bipartite:coherence:on:diagonal} and \eqref{expressions:for:the:first:order:bipartite:coherence:off:diagonal} we faced no conceptual hurdles but only lengthy algebra. We do not provide all the steps of the computations here for the sake of clarity of presentation.

%----------------------------------------------------------------------------------------------------------------------------------------------------------------------------------------------------------------------------%
\subsection{Mixedness of the final reduced state of the mechanical resonators}
%----------------------------------------------------------------------------------------------------------------------------------------------------------------------------------------------------------------------------%
Given our chosen initial state $\hat{\rho}(0)$, and the final reduced state of the mechanical resonators $\hat{\rho}_{\textrm{m}}(t)$, we can compute the mixedness induced by the full evolution $\hat{U}_{\textrm{full}}$. This gives us the linear entropy for finite temperature, which has a simple and analytical expression
\begin{align}\label{applied:full:expression:linear:entropy:main:result}
S_N=&1-\sum_{\overset{\{n_k,m_{k}\}}{k\in\mathcal{I}}} \prod_{k\in\mathcal{I}}\,e^{-2\,|\mu_k|^2}\,\frac{|\mu_k|^{2\,(n_k+m_k)}}{n_k!m_k!}\,\prod_p \frac{\exp\left[-\frac{1}{\cosh (2\,r_p)}\left|\Delta^{(p)}_{\{n_k,m_{k}\}}\right|^2\right]}{\cosh (2\,r_p)}.
\end{align}
We can now look at the contribution when the nonlinearity is switched off, i.e., $\left|\Delta^{(p)}_{\{n_k,m_{k}\}}\right|=0$. This implies that \eqref{applied:full:expression:linear:entropy:main:result} yields the mixedness $S^{\textrm{in}}_{N}$ of the initial state, which simply reads
\begin{align}\label{applied:full:expression:linear:entropy:main:result:in:the:middle}
S^{\textrm{in}}_N=&1-\sum_{\overset{\{n_k,m_{k}\}}{k\in\mathcal{I}}} \prod_{k\in\mathcal{I}}\,e^{-2\,|\mu_k|^2}\,\frac{|\mu_k|^{2\,(n_k+m_k)}}{n_k!m_k!}\,\prod_p \frac{1}{\cosh (2\,r_p)}=1-\prod_p \frac{1}{\cosh (2\,r_p)}.
\end{align}
This allows us to express the full mixedness \eqref{applied:full:expression:linear:entropy:main:result} as 
\begin{align}\label{applied:full:expression:linear:entropy:main:result:final}
S_N=&S^{\textrm{in}}_N+\sum_{\overset{\{n_k,m_{k}\}}{k\in\mathcal{I}}} \prod_{k\in\mathcal{I}}\,e^{-2\,|\mu_k|^2}\,\frac{|\mu_k|^{2\,(n_k+m_k)}}{n_k!m_k!}\, \frac{1-\prod_p\exp\left[-\frac{1}{\cosh (2\,r_p)}\left|\Delta^{(p)}_{\{n_k,m_{k}\}}\right|^2\right]}{\prod_p\cosh (2\,r_p)},
\end{align}
which is one of our main results. 

The zero temperature $T=0$ case is simply obtained by setting $r_p=0$ for all $p$. When the temperature becomes increasingly high, it tends to inhibit the generation of mixedness, i.e., correlations between the two systems. This is in line with previous results that studied the competition between initial mixedness (due to temperature) and the coherent generation of excitations \cite{Bruschi:Friis:2013}.

It is now clear from \eqref{applied:full:expression:linear:entropy:main:result:final} that the last fraction in the expression is the \emph{direct} and \emph{full} contribution of the nonlinearity to the mixedness, since it vanishes (together with the whole expression) for vanishing nonlinear coupling.

%----------------------------------------------------------------------------------------------------------------------------------------------------------------------------------------------------------------------------%
\section{Applications: optomechanics}\label{section:application:to:optomechanics}
%----------------------------------------------------------------------------------------------------------------------------------------------------------------------------------------------------------------------------%
Now that we have obtained our results for a general setup of many cavity modes interacting with many mechanical resonators, we can apply them to simple scenarios that model cases of physical interest. In particular, we focus on Hamiltonians that describe more in particular standard \textit{closed} optomechanical systems without external drive (i.e., we still consider the evolution through the nonlinear unitary operator $\xi_p^{(\pm)}=0$ for all $p$). 

Let us assume that $\lambda_p^{(\pm)}=g^{(-)}_{np}=0$ and call $g^{(+)}_{np}(t)\equiv g_{np}(t)$, that is, we will consider the Hamiltonian \eqref{main:big:time:dependent:Hamiltonian:to:decouple} which now reads
\begin{align}\label{main:applied:Hamiltonian}
\hat {H}_{\textrm{OM}}=& \sum_n \hbar\,\omega_{\textrm{c},n} \hat a_n^\dagger \hat a_n + \sum_p\hbar\,\omega_{\textrm{m},p} \,\hat b_p^\dagger \hat b_p+\sum_{n,p}\hbar\,g_{np} \hat a^\dagger_n\hat a_n \hat{B}^{(+)}_p.
\end{align}
This implies also that $\tilde{F}_{\textrm{c},n}^{(p)}=F_{(\pm)}^{(p)}=0$, and therefore the time evolution operator \eqref{solution:to:the:decoupling:of:full:nonlinear:part} that here reduces to
\begin{align}
\hat{U}_{\textrm{OM}}(t)=&e^{-i\,\sum_n \omega_{\textrm{c},n}\,\hat{N}_n\,t}\,e^{-i\,\sum_p F_{\textrm{m}}^{(p)}\hat{b}^\dag_p \hat{b}_p}\,e^{-\frac{i}{2} \sum_{nmp} F_{nm}^{(p)} \hat{N}_{nm}}\,e^{-i\,\sum_{np} F_n^{(p,+)}\hat{N}_n \hat{B}^{(p,+)}}\, e^{-i\,\sum_{np} F_n^{(p,-)} \hat{N}_n \hat{B}^{(p,-)}}.
\end{align}
Note that, for each fixed oscillator $p$, this coincides exactly with the expression obtained in the literature \cite{Bruschi:Xuereb:2018}.

%----------------------------------------------------------------------------------------------------------------------------------------------------------------------------------------------------------------------------%
\subsection{Linearised regime}
%----------------------------------------------------------------------------------------------------------------------------------------------------------------------------------------------------------------------------%
Our techniques allow, in principle, to fully grasp the nonlinear character of the system. Previously, such analytical understanding was not available and a standard approach has been to \textit{linearise} the standard optomechanical Hamiltonian, i.e., the Hamiltonian \eqref{main:applied:Hamiltonian}. This procedure is standard and we refer to the literature for an extensive justification \cite{Aspelmeyer:Kippenberg:2014}.

In brief, we consider the interaction term $\hbar\,g_{np} \hat a^\dagger_n\hat a_n \hat{B}^{(+)}_p$ and replacing $\hat{a}_n\rightarrow \alpha_n+\delta\hat{a}_n$, where $\alpha_n\gg1$ represents a large mean value for the cavity operators and the ``new'' operator  $\delta\hat{a}_n$ represents small excitations around the large classical mean $\alpha_n$, which we assume real for simplicity. The operator $\delta\hat{a}_n$ has vanishing first moment, i.e., $\langle\delta\hat{a}_n\rangle=0$. In addition, the linearisation procedure requires us to start in an initial state where $\mu_n=0$ for all $n$, that is., all cavity modes are initially in the vacuum. The final step is to keep only terms proportional to $\alpha_n$ and $\alpha_n^2$. 
What is left is called the \textit{linearised Hamiltonian} $\hat {H}_{\textrm{lin}}$, which reads 
\begin{align}\label{main:linearised:time:dependent:Hamiltonian:to:decouple}
\hat {H}_{\textrm{lin}}=& \sum_n \hbar\,\omega_{\textrm{c},n} \delta\hat  a_n^\dagger \delta\hat a_n + \sum_p\hbar\,\omega_{\textrm{m},p} \,\hat b_p^\dagger \hat b_p+\sum_p\hbar\,g_{np}\,\alpha_n^2\,\hat{B}^{(+)}_p+\sum_{n,p}\hbar\,\alpha_n\,g_{np}\,\hat{A}^{(+)}_n\,\hat{B}^{(+)}_p,
\end{align}
where it is evident that the Hamiltonian contains only terms that are at most quadratic in the creation and annihilation operators. In particular, the last term is the central contribution arising from the linearisation of the nonlinear part of the initial Hamiltonian \eqref{main:big:time:dependent:Hamiltonian:to:decouple}. Here we also have $\hat{A}^{(+)}_n:=\delta\hat  a_n^\dagger+ \delta\hat a_n$

The time evolution induced by the linearised Hamiltonian \eqref{main:linearised:time:dependent:Hamiltonian:to:decouple} reads $\hat{U}_{\textrm{lin}}(t)=\hat{U}_0(t)\,\overset{\leftarrow}{\mathcal{T}}\,\exp[-i/\hbar\,\int_0^t\,dt'\,\hat{H}_1]$, where we have introduced $\hat{U}_0(t)=\exp[-i( \sum_n \omega_{\textrm{c},n} \delta\hat  a_n^\dagger \delta\hat a_n + \sum_p\omega_{\textrm{m},p} \,\hat b_p^\dagger \hat b_p)\,t]$ and
\begin{align}\label{main:linearised:time:dependent:Hamiltonian:to:decouple:convenient}
\hat{H}_1=&+\sum_{n,p}\hbar\,\alpha_n\,g_{np}\,\cos((\omega_{\textrm{c},n}+\omega_{\textrm{m},p})\,t)\,\left(\delta\hat{a}^\dag_n\hat{b}^\dag_p+\delta\hat{a}_n\hat{b}_p\right)+\sum_{n,p}\hbar\,\alpha_n\,g_{np}\,\sin((\omega_{\textrm{c},n}+\omega_{\textrm{m},p})\,t)\,i\,\left(\delta\hat{a}^\dag_n\hat{b}^\dag_p-\delta\hat{a}_n\hat{b}_p\right)\nonumber\\
&+\sum_{n,p}\hbar\,\alpha_n\,g_{np}\,\cos((\omega_{\textrm{c},n}-\omega_{\textrm{m},p})\,t)\,\left(\delta\hat{a}^\dag_n\hat{b}_p+\delta\hat{a}_n\hat{b}^\dag_p\right)+\sum_{n,p}\hbar\,\alpha_n\,g_{np}\,\sin((\omega_{\textrm{c},n}-\omega_{\textrm{m},p})\,t)\,i\,\left(\delta\hat{a}^\dag_n\hat{b}_p-\delta\hat{a}_n\hat{b}^\dag_p\right)\nonumber\\
&+\sum_p\hbar\,g_{np}\,\alpha_n^2\,\cos(\omega_{\textrm{m},p}\,t)\,\hat{B}^{(+)}_p+\sum_p\hbar\,g_{np}\,\alpha_n^2\,\sin(\omega_{\textrm{m},p}\,t)\,\hat{B}^{(-)}_p\nonumber\\
\end{align}
Decoupling of the time evolution $\hat{U}_{\textrm{lin}}$ induced by this Hamiltonian can in principle be done analytically \cite{Bruschi:Lee:2013}. It requires $(N+M)\,(2\,(N+M)+1)$ terms for the quadratic part and $2\,M$ for the order one part (i.e., the displacement part of the hamiltonian) in the decomposition \eqref{general:time:evolution:operator:decoupled:general}. Here $N$ and $M$ indicate the number of cavity modes and mechanical modes respectively. 

We do not want to proceed to obtain such solution, since it would require extremely cumbersome algebra which is unnecessary for our purposes here.
Instead, in the following, we will use the full result and the linearised one to discuss the usefulness of the work done here.

%----------------------------------------------------------------------------------------------------------------------------------------------------------------------------------------------------------------------------%
\subsection{The weak coupling regime}
%----------------------------------------------------------------------------------------------------------------------------------------------------------------------------------------------------------------------------%
We have obtained exact expressions for all quantities of interest in this work. We can now look at a regime of weak coupling between light and matter, that is, to assume that the couplings $g_{np}$ in the Hamiltonian \eqref{main:applied:Hamiltonian} are small, i.e., $g_{np}(t)=\epsilon\,\tilde{g}_{np}(t)$, where $\epsilon\ll1$ and $\tilde{g}_{np}(t)$ is some time-dependent function that remains finite at all times, such that $g_{np}^{(+)}(t)\ll1$. We assume that the parameter $\epsilon$ is the same for all couplings for the sake of simplicity.

%----------------------------------------------------------------------------------------------------------------------------------------------------------------------------------------------------------------------------%
\subsubsection{The weak coupling regime: bipartite coherence}
%----------------------------------------------------------------------------------------------------------------------------------------------------------------------------------------------------------------------------%
We now compute the bipartite coherence induced by the full nonlinear Hamiltonian and by the linearised Hamiltonian \eqref{main:linearised:time:dependent:Hamiltonian:to:decouple:convenient}.

For the full nonlinear setup we have
\begin{align}\label{first:order:bipartite:coherence:application:weak:coupling}
g^{(1)}_{\tilde{k}\tilde{k}'}(t)\sim&1+\mathcal{O}(\epsilon^2)\nonumber\\
g^{(1)}_{\tilde{k}p'}(t)\sim&\frac{1}{\sqrt{N_{\textrm{i},p'}}}\left|N_{\textrm{i},p'}\,\int_0^t\,dt'\,\tilde{g}_{\tilde{k}p'}(t)\,e^{-i\,\omega_{\textrm{m},p'}\,t'}-\sum_{n\in\mathcal{I}} |\mu_n|^2\,\int_0^t\,dt'\,\tilde{g}_{np'}(t)\,e^{-i\,\omega_{\textrm{m},p'}\,t'}\right|\,\epsilon\nonumber\\
g^{(1)}_{pp'}(t)\sim&\mathcal{O}(\epsilon^2),
\end{align}
while, in the linearised regime we find
\begin{align}\label{first:order:bipartite:coherence:linearised:application:weak:coupling}
g^{(1)}_{\tilde{k}\tilde{k}'}(t)\sim&1+\mathcal{O}(\epsilon^2)\nonumber\\
g^{(1)}_{\tilde{k}p'}(t)\sim&\sqrt{N_{\textrm{i},p'}}\left|\int_0^t\,dt'\,\tilde{g}_{\tilde{k}p'}(t)\,e^{-i\,(\omega_{\textrm{c},\tilde{k}}+\omega_{\textrm{m},p'})\,t'}\right|\,\epsilon\nonumber\\
g^{(1)}_{pp'}(t)\sim&\mathcal{O}(\epsilon^2)
\end{align}
These two expressions are different. If coherence could be measured experimentally, we should find agreement with the expression obtained through the full nonlinear Hamiltonian.

%----------------------------------------------------------------------------------------------------------------------------------------------------------------------------------------------------------------------------%
\subsubsection{The weak coupling regime: mixedness of the mechanical resonator subsystem}
%----------------------------------------------------------------------------------------------------------------------------------------------------------------------------------------------------------------------------%
We can now ask the following question: how much mixedness is induced if the nonlinear coupling is weak? 

It is not difficult to show, using perturbation theory, that the linear entropy $S_N$ for the full nonlinear case and the linearized case reads
\begin{align}\label{applied:full:expression:linear:entropy:approximate:result:final}
S_N\sim&S^{\textrm{in}}_N+\mathcal{O}(\epsilon^2).
\end{align}
This implies that, to the order we are interested in, the optical and mechanical subsystem remain separable. 

We note here that there is coherence induced between the optical modes and the mechanical ones at first order in $\epsilon$ in this regime, as seen in \eqref{first:order:bipartite:coherence:application:weak:coupling} and \eqref{first:order:bipartite:coherence:linearised:application:weak:coupling}. However, the mixedness induced between the states of the subsystems occurs only to second order in $\epsilon$. This is not an unknown feature of such properties of quantum states in perturbation theory \cite{Sabin:Kohlrus:2016}.

%----------------------------------------------------------------------------------------------------------------------------------------------------------------------------------------------------------------------------%
\subsection{Applications to optomechanics: single mode \& multi-resonator cavity}
%----------------------------------------------------------------------------------------------------------------------------------------------------------------------------------------------------------------------------%
We specialise to standard optomechanical scenarios with one cavity mode $\hat{a}_{\tilde{k}}$ and an arbitrary number of mechanical resonators $\hat{b}_p$. This means that the set $\mathcal{I}=\{\tilde{k}\}$, which has one element.

The first quantities we can compute are the time evolution of the mode operators. Using the results obtained above we find
\begin{align}\label{time:evolution:of:basic:operators:optomechanical:case}
	\hat{a}_{\tilde{k}}(t) &= e^{-i \left[\omega_{\tilde{k}} t+\sum_p\,F^{(p)}_{\tilde{k}\tilde{k}}+\sum_{n,p}\left(F^{(p)}_{\{\tilde{k}n\}}+2 F^{(p,+)}_{\tilde{k}}\,F^{(p,-)}_n\right)\hat{a}^\dag_n\hat{a}_n\right]} e^{-i\,\sum_p\,F^{(p,+)}_{\tilde{k}}\,\hat{B}^{(p)}_+} \hat{a}_n\nonumber\\
	\hat{b}_p(t) &= e^{-i\,F_{\textrm{m}}^{(p)}}\,\left[\hat{b}_p-i\,F_{\tilde{k}}^{(p)}\,\hat{a}^\dag_{\tilde{k}}\hat{a}_{\tilde{k}}\right].
\end{align}
It is then easy to check that, given our chosen initial state, we have $\langle \hat{a}_{\tilde{k}}^\dag(t)\,\hat{a}_{\tilde{k}}(t)\rangle=\langle \hat{a}_{\tilde{k}}^\dag(0)\,\hat{a}_{\tilde{k}}(0)\rangle$ and $\langle \hat{b}_p^\dag(t)\,\hat{b}_p(t)\rangle=N_{\mathrm{i},p}+|F_{\tilde{k}}^{(p)}|^2\,(\mu_{\tilde{k}}^2+\mu_{\tilde{k}}^4)$.

We can also compute the first-order bipartite quantum coherence \eqref{first-order:amplitude:correlation:function} for the field mode and an oscillator which reads
\begin{align}\label{first:order:bipartite:coherence:application:field:mode:many:resonators}
g^{(1)}_{\tilde{k}p'}(t)=&\frac{|F_{\tilde{k}}^{(p')}|\,\left|N_{\mathrm{i},p'}-e^{2\,i\,\phi_{\tilde{k}}}\,|\mu_{\tilde{k}}|^2\right|}{\sqrt{N_{\mathrm{i},p'}+|F_{\tilde{k}}^{(p')}|^2\,|\mu_{\tilde{k}}|^2\,(1+|\mu_{\tilde{k}}|^2)}}\,e^{-2\,|\mu_{\tilde{k}}|^2\,\sin^2\,\phi_{\tilde{k}}}\,e^{-\frac{1}{2}\,\sum_p\,(1+2\,N_{\mathrm{i},p})\,|F_{\tilde{k}}^{(p)}|^2}
\end{align}
and for two oscillators, which gives us
\begin{align}\label{first:order:bipartite:coherence:application:many:resonators:many:resonators}
g^{(1)}_{pp'}(t)=&\frac{|F_{\tilde{k}}^{(p)}|\,|F_{\tilde{k}}^{(p')}|\,|\mu_{\tilde{k}}|^2\,(1+|\mu_{\tilde{k}}|^2)}{\sqrt{N_{\mathrm{i},p}+|F_{\tilde{k}}^{(p)}|^2\,|\mu_{\tilde{k}}|^2\,(1+|\mu_{\tilde{k}}|^2)}\,\sqrt{N_{\mathrm{i},p'}+|F_{\tilde{k}}^{(p')}|^2\,|\mu_{\tilde{k}}|^2\,(1+|\mu_{\tilde{k}}|^2)}}
\end{align}
Above, we have defined the angle $\phi_{\tilde{k}}:=\frac{1}{2}\,\sum_p\,(F^{(p)}_{\{\tilde{k}\tilde{k}\}}+2\,F^{(p,+)}_{\tilde{k}}\,F^{(p,-)}_{\tilde{k}})$. Note that, for zero temperature we have $r_p=0$ for all $p$, and the expressions \eqref{first:order:bipartite:coherence:application:field:mode:many:resonators} and \eqref{first:order:bipartite:coherence:application:many:resonators:many:resonators} simplify and reduce to 
\begin{align}
g^{(1)}_{\tilde{k}p'}(t)=&\frac{|\mu_{\tilde{k}}|}{\sqrt{1+|\mu_{\tilde{k}}|^2}}\,e^{-2\,|\mu_{\tilde{k}}|^2\,\sin^2\,\phi_{\tilde{k}}}\,e^{-\frac{1}{2}\,\sum_p\,|F_{\tilde{k}}^{(p)}|^2}
\end{align}
and $g^{(1)}_{pp'}(t)=1$. This means that, while the mode of light and any resonator mode are coherent with a strength that depends on the parameters of the problem, any pair of resonators is perfectly coherent at zero temperature. We also note that, in this regime, the light and resonator coherence decreases exponentially with $|\mu_{\tilde{k}}|$, unless $\phi_{\tilde{k}}=0$. In the limit $|\mu_{\tilde{k}}|\rightarrow\infty$ we have that $g^{(1)}_{\tilde{k}p'}(t)=\exp[-\frac{1}{2}\,\sum_p\,|F_{\tilde{k}}^{(p)}|^2]\neq0$ only at the times $t_n$ such that $\phi_{\tilde{k}}(t_n)=0$ and vanishes for all other times. This implies that, to be able to verify the coherence between light and a single resonator it is necessary to reduce the number of photons in the coherent state as much as possible.

For finite temperature the issue becomes more delicate. It is clear from \eqref{first:order:bipartite:coherence:application:field:mode:many:resonators} and \eqref{first:order:bipartite:coherence:application:many:resonators:many:resonators} that an increase in temperature (i.e., an increase in $r_p$ for all $p$) implies that it is more difficult to establish the desired coherence. Therefore, reducing the temperature to levels where the initial phononic population $N_{\mathrm{i},p}$ fore each resonator becomes small is paramount.

We can also compute the mixedness of the reduced state of the oscillators, which employs algebraic manipulations that can be found in \ref{summing:mixedness:special:case:one:mode:many:resonators:subsection}. Finally, we show that it reads
\begin{align}\label{applied:full:expression:linear:entropy:specific:result:final}
S_N=&S^{\textrm{in}}_N+2\,e^{-2\,|\mu_{\tilde{k}}|^2}\,\frac{\sum_{m=1}^{+\infty}\,I_m(2\,|\mu_{\tilde{k}}|^2)\,\left(1-e^{-\sum_p\frac{1}{(1+2\,N_{\mathrm{i},p})}\,\left|F^{(p)}_{\tilde{k}}\right|^2\,m^2}\right)}{\prod_p\,(1+2\,N_{\mathrm{i},p})}.
\end{align}

%\begin{align}\label{applied:full:expression:linear:entropy:specific:result:final}
%S_N=&S^{\textrm{in}}_N+\frac{1-e^{-2\,|\mu_{\tilde{k}}|^2}\,\left[I_0(2\,|\mu_{\tilde{k}}|^2)+2\,\sum_{d=1}^{+\infty}\,I_d(2\,|\mu_{\tilde{k}}|^2)\,e^{-\sum_p\frac{1}{\cosh (2\,r_p)}\,\left|F^{(p)}_{\tilde{k}}\right|^2\,d^2}\right]}{\prod_p\cosh (2\,r_p)}.
%\end{align}
In this formula we have introduced the modified Bessel functions $I_n(z)$. Notice that, when $F^{(p)}_{\tilde{k}}=0$ we recover immediately $S_N=S^{\textrm{in}}_N$ as expected.

%----------------------------------------------------------------------------------------------------------------------------------------------------------------------------------------------------------------------------%
\subsection{Modulated scenario for a single mode \& multi-resonator cavity}
%----------------------------------------------------------------------------------------------------------------------------------------------------------------------------------------------------------------------------%
We can now ask another question of potential physical interest, which can also be used to highlight the difference in the predictions obtained with the full time-evolution obtained here, and with the linearised approach.
We consider a scenario where the couplings have a time dependence of the form $g_{\tilde{k}p}(t)=g_{\tilde{k}p}\,(1+\kappa\,\sin(\omega_\mathrm{d}\,t))$ or $g_{\tilde{k}p}(t)=g_{\tilde{k}p}\,(1+\kappa\,\cos(\omega_\mathrm{d}\,t))$ for some modulating frequency $\omega_\mathrm{d}$ and amplitude of the oscillation $\kappa$.

This form of the coupling can be used to model in a simple way an interaction strength between light and matter that is not constant, something that is typically assumed as the lowest order approximation in many experiments. However, no quantity of this kind can be fundamentally and ultimately constant. Therefore, the expression that we provide here is provides a reasonable toy model to study deviations from the ideal case, where the mean value $g_{\tilde{k}p}$ is the one measured in the laboratory. This form of coupling has already been proposed and used in for different studies \cite{Qvarfort:Serafini:2018}.

Here we choose to look at two possible situations: $\omega_\mathrm{d}=\omega_{\textrm{c},\tilde{k}}\pm\omega_{\textrm{m},\tilde{p}}$. Here $\tilde{k}$ and $\tilde{p}$ are a specific optical and mechanical mode respectively. In \ref{linearised:driven:section} we show that, depending on the choice, in the linearised scenario this corresponds to a two-mode squeezing operation (+ sign) or mode-mixing operation (- sign) after a sufficiently long time (i.e., an approximation that is also used in the rotating wave approximation). Here we do not discuss the details of how much time needs to pass for this approximation to become more and more accurate but we just note that such resonant regime naturally occurs across all of physics and we leave it to future work to study the details of potential implementations.

%----------------------------------------------------------------------------------------------------------------------------------------------------------------------------------------------------------------------------%
\subsection{Modulated scenario: linearised results}
%----------------------------------------------------------------------------------------------------------------------------------------------------------------------------------------------------------------------------%
Let us start with the linearised case.
When $\omega_\mathrm{d}=\omega_{\textrm{c},\tilde{k}}+\omega_{\textrm{m},\tilde{p}}$  we find that the evolution of the operators $\delta\hat{a}_{\tilde{k}}$ and $\hat{b}_{\tilde{p}}$ in the linearised regime after sufficiently long time reads
\begin{align}\label{main:linearised:time:dependent:Hamiltonian:to:decouple:convenient:squeezing}
\hat{U}_{\textrm{lin}}(t)\,\delta\hat{a}_{\tilde{k}}\,\hat{U}_{\textrm{lin}}(t)\sim&e^{-i\,\omega_{\textrm{c},\tilde{k}}\,t}\left[\delta\hat{a}_{\tilde{k}}\,\cosh(\alpha_{\tilde{k}}\,\kappa\,g_{\tilde{k}\tilde{p}}\,t)+\hat{b}_{\tilde{p}}^\dag\,\sinh(\alpha_{\tilde{k}}\,\kappa\,g_{\tilde{k}\tilde{p}}\,t)\right]\nonumber\\
\hat{U}_{\textrm{lin}}^\dag(t)\,\hat{b}_{\tilde{p}}\,\hat{U}_{\textrm{lin}}(t)\sim&e^{-i\,\omega_{\textrm{m},\tilde{p}}\,t}\left[\hat{b}_{\tilde{p}}\,\cosh(\alpha_{\tilde{k}}\,\kappa\,g_{\tilde{k}\tilde{p}}\,t)+\delta\hat{a}_{\tilde{k}}^\dag\,\sinh(\alpha_{\tilde{k}}\,\kappa\,g_{\tilde{k}\tilde{p}}\,t)\right]
\end{align}
while, when $\omega_\mathrm{d}=\omega_{\textrm{c},\tilde{k}}-\omega_{\textrm{m},\tilde{p}}$ we have, after sufficiently long time, 
\begin{align}\label{main:linearised:time:dependent:Hamiltonian:to:decouple:convenient:modemixing}
\hat{U}_{\textrm{lin}}(t)\,\delta\hat{a}_{\tilde{k}}\,\hat{U}_{\textrm{lin}}(t)=&e^{-i\,\omega_{\textrm{c},\tilde{k}}\,t}\left[\delta\hat{a}_{\tilde{k}}\,\cos(\alpha_{\tilde{k}}\,\kappa\,g_{\tilde{k}\tilde{p}}\,t)+\hat{b}_{\tilde{p}}\,\sin(\alpha_{\tilde{k}}\,\kappa\,g_{\tilde{k}\tilde{p}}\,t)\right]\nonumber\\
\hat{U}_{\textrm{lin}}^\dag(t)\,\hat{b}_{\tilde{p}}\,\hat{U}_{\textrm{lin}}(t)=&e^{-i\,\omega_{\textrm{m},\tilde{p}}\,t}\left[\hat{b}_{\tilde{p}}\,\cos(\alpha_{\tilde{k}}\,\kappa\,g_{\tilde{k}\tilde{p}}\,t)-\delta\hat{a}_{\tilde{k}}\,\sinh(\alpha_{\tilde{k}}\,\kappa\,g_{\tilde{k}\tilde{p}}\,t)\right].
\end{align}
These expressions allow us to compute the expectation values of the population operators in these regimes. Recalling that we defined $\hat{a}_k=\alpha_k+\delta\hat{a}_k$, for $\omega_\mathrm{d}=\omega_{\textrm{c},\tilde{k}}+\omega_{\textrm{m},\tilde{p}}$ we have 
\begin{align}\label{expectation:values:linearised:driven:squeezed}
\langle\hat{a}_{\tilde{k}}^\dag(t)\hat{a}_{\tilde{k}}(t)\rangle\sim&|\alpha_{\tilde{k}}|^2+N_{\textrm{i},\tilde{p}}\,\sinh^2(\alpha_{\tilde{k}}\,\kappa\,g_{\tilde{k}\tilde{p}}\,t)\nonumber\\
\langle\hat{b}_{\tilde{p}}^\dag(t)\hat{b}_{\tilde{p}}(t)\rangle\sim&N_{\textrm{i},\tilde{p}}\,\left(1+\sinh^2(\alpha_{\tilde{k}}\,\kappa\,g_{\tilde{k}\tilde{p}}\,t)\right),
\end{align}
while for $\omega_\mathrm{d}=\omega_{\textrm{c},\tilde{k}}-\omega_{\textrm{m},\tilde{p}}$ we have 
\begin{align}\label{expectation:values:linearised:driven:modemixed}
\langle\hat{a}_{\tilde{k}}^\dag(t)\hat{a}_{\tilde{k}}(t)\rangle\sim&|\alpha_{\tilde{k}}|^2+N_{\textrm{i},\tilde{p}}\,\sin^2(\alpha_{\tilde{k}}\,\kappa\,g_{\tilde{k}\tilde{p}}\,t)\nonumber\\
\langle\hat{b}_{\tilde{p}}^\dag(t)\hat{b}_{\tilde{p}}(t)\rangle\sim&N_{\textrm{i},\tilde{p}}\,\left(1-\sin^2(\alpha_{\tilde{k}}\,\kappa\,g_{\tilde{k}\tilde{p}}\,t)\right),
\end{align}
Notice that both the resonant and asymptotic (i.e., after a sufficiently long time) behaviors \eqref{expectation:values:linearised:driven:squeezed} and \eqref{expectation:values:linearised:driven:modemixed} must be compatible with the fact that we have linearised the system, that is, $\delta\hat{a}$ is an operator that creates excitations around the vacuum. Therefore, we believe the validity of these expressions for ``large'' times, which do not invalidate the linearisation procedure. The exact details are left for future work.

%----------------------------------------------------------------------------------------------------------------------------------------------------------------------------------------------------------------------------%
\subsection{Modulated scenario: full results}
%----------------------------------------------------------------------------------------------------------------------------------------------------------------------------------------------------------------------------%
In the case of the results obtained in this work, we note immediately through the expressions \eqref{solution:to:major:differential:equations:with:all:resonators} and \eqref{time:evolution:of:basic:operators:optomechanical:case} that we \textit{do not} have any resonant behavior when $\omega_\mathrm{d}=\omega_{\textrm{c},\tilde{k}}\pm\omega_{\textrm{m},\tilde{p}}$. In fact, the expressions for the time evolution operator \eqref{time:evolution:of:basic:operators:optomechanical:case} imply that any resonance would occur for $\omega_\mathrm{d}=\omega_{\textrm{m},\tilde{p}}$, which is \textit{in stark contrast} with the results obtained above and displayed in \eqref{expectation:values:linearised:driven:squeezed} and \eqref{expectation:values:linearised:driven:modemixed}. 

In particular, we obtain
\begin{align}\label{expectation:values:linearised:driven:full}
\langle\hat{a}_{\tilde{k}}^\dag(t)\hat{a}_{\tilde{k}}(t)\rangle=&|\mu_{\tilde{k}}|^2\nonumber\\
\langle\hat{b}_{\tilde{p}}^\dag(t)\hat{b}_{\tilde{p}}(t)\rangle=&N_{\textrm{i},\tilde{p}}+g^2_{\tilde{k}p}\,\left|\int_0^t dt'\,(1+\kappa\,\sin(\omega_\mathrm{d}\,t'))\,e^{-i\,\omega_{\mathrm{m},p} t'}\right|\,\left(|\mu_{\tilde{k}}|^2+|\mu_{\tilde{k}}|^4\right).
\end{align} 
If we drive the system $\omega_\mathrm{d}=\omega_{\textrm{m},\tilde{p}}$, after a sufficiently long time we would obtain
\begin{align}\label{expectation:values:linearised:driven:full:resonant}
\langle\hat{a}_{\tilde{k}}^\dag(t)\hat{a}_{\tilde{k}}(t)\rangle=&|\mu_{\tilde{k}}|^2\nonumber\\
\langle\hat{b}_{\tilde{p}}^\dag(t)\hat{b}_{\tilde{p}}(t)\rangle\sim&N_{\textrm{i},\tilde{p}}+\frac{1}{4}\,g^2_{\tilde{k}\tilde{p}}\,\kappa^2\,\left(|\mu_{\tilde{k}}|^2+|\mu_{\tilde{k}}|^4\right)\,t^2.
\end{align} 

%----------------------------------------------------------------------------------------------------------------------------------------------------------------------------------------------------------------------------%
\subsection{Modulated scenario: considerations}
%----------------------------------------------------------------------------------------------------------------------------------------------------------------------------------------------------------------------------%
In any realistic implementation one has $\omega_{\textrm{c},\tilde{k}}\gg\omega_{\textrm{m},\tilde{p}}$, and therefore we see that the the combination $\omega_{\textrm{c},\tilde{k}}\pm\omega_{\textrm{m},\tilde{p}}$ would never take values very close to $\omega_{\textrm{m},\tilde{p}}$. 

For these reasons, the resonant regimes can be distinguished through experimental efforts, by driving the system at completely different drive frequencies. Our results show that, by driving the system with a coupling of the form $g\propto(1+\kappa\,\sin(\omega_\mathrm{d}\,t))$, where the drive frequency matches that of one of the mechanical resonators, we should observe a time dependent population of phonons predicted by \eqref{expectation:values:linearised:driven:full:resonant}.  By driving the system at the frequency $\omega_\mathrm{d}=\omega_{\textrm{c},\tilde{k}}-\omega_{\textrm{m},\tilde{p}}$, we should \textit{not} observe any resonant behavior, contrary to what predicted by the linearised regime in \eqref{expectation:values:linearised:driven:squeezed}.

%----------------------------------------------------------------------------------------------------------------------------------------------------------------------------------------------------------------------------%
\section{Conclusion}\label{conclusions}
%----------------------------------------------------------------------------------------------------------------------------------------------------------------------------------------------------------------------------%
In this work we found an analytic expression for the time evolution of an arbitrary number of coupled bosonic modes interacting through a time-dependent optomechanical-like Hamiltonian. Despite of the system having potentially an arbitrarily large amount of constituents, and despite of the nonlinearity present in the interaction driven by a time-dependent coupling, we were able to decouple the time-evolution operator using tools developed for this purpose \cite{Bruschi:Lee:2013,Bruschi:Xuereb:2018}. This result does not rely on any approximation and is therefore general: the only assumption made is that the Hamiltonian has the form \eqref{main:big:time:dependent:Hamiltonian:to:decouple}, which occurs in many systems, such as optomechanical cavities \cite{Aspelmeyer:Kippenberg:2014}.

We were able to compute the time evolution of the expectation values of meaningful operators, such as the average photonic and phononic excitation, the first-order quantum bipartite coherence, and the mixedness of the reduced state of the mechanical oscillators. Our results allow us to study the coherence induced between the subsystems due to the nonlinear interaction of light and matter. Furthermore, they allow us to clearly quantify, using the linear entropy, the increase of mixedness of the subsystem of the resonators when the nonlinearity is switched on, as a function of time, the initial photonic population, and temperature inside the cavity. In addition, we compared some of the predictions with those arising from the linearised version of our system. Linearisation is typically used to describe optomechanical systems, and we have found that there are qualitative differences that arise when quantifying, for example, the expectation value of the cavity mode population or of the phononic population of the mechanical excitations. We gave a concrete example where the light matter coupling can be modulated periodically, and we were able to show that the linearised model and the full nonlinear one treated here, give different results. 

Given the lack of a systematic understanding of the nonlinear nature of the interaction, and the nonlinearity induced by the coupling, this analytical insight can help shedding light onto intrinsic nonlinear aspects of quantum (opto) mechanical systems. In the end, these insights can also help in the quest of demonstrating in the laboratory the quantum nature of ``macroscopic objects'', such as the mechanical resonators. 
Finally, the decoupling obtained here can be applied to many situations of theoretical and practical interest. We leave it to future work to pursue such new directions.

%----------------------------------------------------------------------------------------------------------------------------------------------------------------------------------------------------------------------------%
\section*{Acknowledgments}
%----------------------------------------------------------------------------------------------------------------------------------------------------------------------------------------------------------------------------%
We acknowledge Alessio Serafini and Andr\'e Xuereb for useful comments and discussions.

\bibliography{OMNLBib}

\newpage

%----------------------------------------------------------------------------------------------------------------------------------------------------------------------------------------------------------------------------%
\appendix
%----------------------------------------------------------------------------------------------------------------------------------------------------------------------------------------------------------------------------%

%----------------------------------------------------------------------------------------------------------------------------------------------------------------------------------------------------------------------------%
\section{Decoupling of the Hamiltonian}\label{Hamitlonian:decoupling:appendix}
%----------------------------------------------------------------------------------------------------------------------------------------------------------------------------------------------------------------------------%
Here we show how to decouple the time-evolution operator $\hat{U}$ induced by the Hamiltonian \eqref{main:time:dependent:Hamiltonian:to:decouple}, reprented here
\begin{align}\label{main:time:dependent:Hamiltonian:to:decouple:appendix}
	\hat {H}=& \sum_n \hbar\,\omega_{\textrm{c},n} \hat a_n^\dagger \hat a_n + \hbar\,\omega_{\textrm{m}} \,\hat b^\dagger \hat b+\hbar\,\lambda^{(+)}\,\hat{B}^{(+)}+ \hbar\,\lambda^{(-)}\,\hat{B}^{(-)}+\sum_{n}\hbar\,g_{n}^{(+)} \hat a^\dagger_n\hat a_n \hat{B}^{(+)}+\sum_{n}\hbar\,g_{n}^{(-)} \hat a^\dagger_n\hat a_n \,\hat{B}^{(-)}.
\end{align}
The procedure to be followed has been developed in the literature \cite{Bruschi:Lee:2013}. All details can be found there.
The time-evolution operator is defined by $\hat {U}:=\overset{\leftarrow}{\mathcal{T}}\,\exp[-i\,\int_0^t\,dt'\,\hat {H}(t')]$.

As discussed in the main text, we make the decoupling ansatz
\begin{align}\label{formal:solution:to:the:decoupling:appendix}
\hat{U}(t)=&e^{-i\,\sum_n F_n\,\hat{N}_n}\,e^{-i\,F_b\,\hat{N}_b}\,e^{-\frac{i}{2}\,\sum_{nm}\,F_{nm}\,\hat{N}_{nm}}\,e^{-i\,F_+\,\hat{B}^{(+)}}\,e^{-i\,\sum_n\,F_n^{(+)}\,\hat{N}_n\,\hat{B}^{(+)}}\,e^{-i\,F_-\,\hat{B}^{(-)}}\,e^{-i\,\sum_n\,F_n^{(-)}\,\hat{N}_n\,\hat{B}^{(-)}},
\end{align}
where we have defined the Lie-algebra basis operators
\begin{align}\label{basis:operator:Lie:algebra}
	\hat{N}_n &:= \hat a^\dagger_n \hat a_n &
	\hat{N}_b &:= \hat b^\dagger \hat b 
	& \hat{N}_{nm} &:= \hat a^\dagger_n \hat a_n \hat a^\dagger_m \hat a_m
	\nonumber\\
	\hat{B}^{(+)} &:=  \hat b^\dagger +\hat b &
	\hat{B}^{(-)} &:= i\,(\hat b^\dagger -\hat b) & &
	 & \nonumber\\
	\hat{N}_n\,\hat{B}^{(+)} &:= \hat{N}_n\,(\hat b^{\dagger}+\hat b) &
	\hat{N}_n\,\hat{B}^{(-)} &:= \hat{N}_n\,i\,(\hat b^{\dagger}-\hat b). & &
\end{align}
We now take the time derivative on both sides of the expression \eqref{formal:solution:to:the:decoupling:appendix} and then multiply on the right by $\hat{U}^\dag(t)$ to obtain
\begin{align}\label{similarity:hamiltonian:appendix}
\frac{1}{\hbar}\hat{H}=&\sum_n \dot{F}_n\,\hat{N}_n+\dot{F}_b\,e^{-i\,\sum_n F_n\,\hat{N}_n}\,\hat{N}_b\,e^{i\,\sum_n F_n\,\hat{N}_n}+\sum_{nm}\dot{F}_{nm}\,e^{-i\,\sum_n F_n\,\hat{N}_n}\,e^{-i\, F_b\,\hat{N}_b}\,\hat{N}_{nm}\,e^{i\, F_b\,\hat{N}_b}\,e^{i\,\sum_n F_n\,\hat{N}_n}+\ldots\nonumber\\
\end{align}
We then use similarity relations of the form 
\begin{align}\label{similarity:relations:appendix}
e^{i\,x\,(\hat{b}^{\dag}+\hat{b})}\,\hat{b}^\dag\,\hat{b}\,e^{-i\,x\,(\hat{b}^{\dag}+\hat{b})}=&\hat{b}^\dag\,\hat{b}-i\,(\hat{b}^{\dag}-\hat{b})\,x+x^2\,\mathds{1};\nonumber\\
e^{x\,(\hat{b}^{\dag}-\hat{b})}\,\hat{b}^\dag\,\hat{b}\,e^{-x\,(\hat{b}^{\dag}-\hat{b})}=&\hat{b}^\dag\,\hat{b}-(\hat{b}^{\dag}+\hat{b})\,x+x^2\,\mathds{1};\nonumber\\
e^{i\,x\,(\hat{b}^{\dag}+\hat{b})}\,i\,(\hat{b}^{\dag}-\hat{b})\,e^{-i\,x\,(\hat{b}^{\dag}+\hat{b})}=&i\,(\hat{b}^{\dag}-\hat{b})-2\,x\,\mathds{1};\nonumber\\
e^{x\,(\hat{b}^{\dag}-\hat{b})}\,(\hat{b}^{\dag}+\hat{b})\,e^{-x\,(\hat{b}^{\dag}-\hat{b})}=&(\hat{b}^{\dag}+\hat{b})-2\,x\,\mathds{1}.
\end{align}
to obtain an explicit expression for \eqref{similarity:hamiltonian:appendix} and, equating coefficients on both sides we obtain the differential equations for the time dependent $F$-functions in terms of the coefficients of the Hamiltonian, which read 
\begin{align}\label{major:differential:equations:appendix}
\omega_\textrm{m}&=\dot{F}_b
\nonumber\\
\omega_{\textrm{c},n}&=\dot{F}_n-2\,\dot{F}_-\,F_n^{(+)}-2\,F_+\,\dot{F}_n^{(-)}
\nonumber\\
0&= \frac{1}{2}\,\dot{F}_{nm}-2\,F_n^{(+)}\,\dot{F}_m^{(-)}
\nonumber\\
\lambda^{(+)} &= \dot{F}_+\,\cos(\omega_b\,t)-\dot{F}_-\,\sin(\omega_b\,t)
\nonumber\\
\lambda^{(-)} &= -\dot{F}_+\,\sin(\omega_b\,t)-\dot{F}_-\,\cos(\omega_b\,t)
\nonumber\\
g_n^{(+)} &=\dot{F}_n^{(+)}\,\cos(\omega_b\,t)-\dot{F}_n^{(-)}\,\sin(\omega_b\,t)
\nonumber\\
g_n^{(-)} &= -\dot{F}_n^{(+)}\,\sin(\omega_b\,t)-\dot{F}_n^{(-)}\,\cos(\omega_b\,t).
\end{align}
Note that the similarity relations \eqref{similarity:relations:appendix} are \emph{not} exhaustive nor complete. They are presented to give an intuitive understanding of the type of relations needed to simplify the expression \eqref{similarity:hamiltonian:appendix}. A complete list can be found in \cite{Bruschi:Xuereb:2018}.

\newpage

%----------------------------------------------------------------------------------------------------------------------------------------------------------------------------------------------------------------------------%
\section{Some useful expressions}\label{useful:expressions}
%----------------------------------------------------------------------------------------------------------------------------------------------------------------------------------------------------------------------------%
In this section we provide some useful expression used in this work.

Let us start by the first, defined as $I_\alpha:=\langle\mu|\exp[-i\,\alpha\,\hat{a}^\dag\,\hat{a}]\,|\mu\rangle$, where $\hat{a}|\mu\rangle=\alpha\,|\mu\rangle$.
We have
\begin{align}\label{first:useful:expression:appendix}
I_\alpha=&\langle\mu|\exp[-i\,\alpha\,\hat{a}^\dag\,\hat{a}]\,|\mu\rangle\nonumber\\
=&\sum_{p,q=0}^{+\infty}e^{-|\mu|^2}\,\frac{(\mu^*)^p\,\mu^q}{\sqrt{p!}\,\sqrt{q!}}\langle p|e^{-i\,\alpha\,\hat{a}^\dag\,\hat{a}}|q\rangle=\sum_{p,q=0}^{+\infty}e^{-|\mu|^2}\,\frac{(\mu^*)^p\,\mu^q}{\sqrt{p!}\,\sqrt{q!}}\,e^{-i\,\alpha\,q}\langle p|q\rangle\nonumber\\
=&\sum_{p=0}^{+\infty}e^{-|\mu|^2}\,\frac{|\mu|^{2\,p}}{p!}\,e^{-i\,\alpha\,p}\nonumber\\
I_\alpha=&e^{-|\mu|^2\,\left(1-e^{-i\,\alpha}\right)}.
\end{align}
Notice that the expression $I^{(n)}_\alpha:=\langle\mu|\,(\hat{a}^\dag\,\hat{a})^n\,\exp[-i\,\alpha\,\hat{a}^\dag\,\hat{a}]\,|\mu\rangle$ can be computed as $\tilde{I}_\alpha=i^n\,\frac{d^n}{d\alpha^n}I_\alpha$. Therefore
\begin{align}\label{second:useful:expression:appendix}
I^{(1)}_\alpha=&|\mu|^2\,e^{-i\,\alpha}\,e^{-|\mu|^2\,\left(1-e^{-i\,\alpha}\right)}.
\end{align}

We continue by computing the expression $J_\alpha:=\sum_{n=0}^{+\infty}T^{2n}/C^2\langle n|\exp[\alpha\,\hat{a}^\dag-\alpha^*\,\hat{a}]\,|n\rangle$, where $T:=\tanh r$ and $C:=\cosh r$.
We have
\begin{align}\label{calculation:one:ok}
J_\alpha=&\sum_{n=0}^{+\infty}\frac{T^{2\,n}}{C^2}\langle n|e^{\alpha\,\hat{a}^\dag-\alpha^*\,\hat{a}}\,|n\rangle\nonumber\\
=&e^{\frac{1}{2}\,|\alpha|^2}\sum_{n=0}^{+\infty}\frac{T^{2\,n}}{C^2}\langle n|e^{-\alpha^*\,\hat{a}}\,e^{\alpha\,\hat{a}^\dag}\,|n\rangle\nonumber\\
=&e^{\frac{1}{2}\,|\alpha|^2}\sum_{n,p,q=0}^{+\infty}\frac{T^{2\,n}}{C^2}\frac{(-\alpha^*)^p\,\alpha^q}{p!\,q!}\langle n|\hat{a}^p\,\hat{a}^{\dag q}\,|n\rangle\nonumber\\
=&e^{\frac{1}{2}\,|\alpha|^2}\sum_{n,p,q=0}^{+\infty}\frac{T^{2\,n}}{C^2}\frac{(-\alpha^*)^p\,\alpha^q}{p!\,q!}\,\sqrt{\frac{(p+n)!}{n!}}\,\sqrt{\frac{(q+n)!}{n!}}\,\langle n+p|n+q\rangle\nonumber\\
=&e^{\frac{1}{2}\,|\alpha|^2}\sum_{n,p=0}^{+\infty}\frac{T^{2\,n}}{C^2}\frac{(-|\alpha|^2)^p}{p!\,p!}\,\frac{(p+n)!}{n!}\nonumber\\
=&e^{\frac{1}{2}\,|\alpha|^2}\sum_{n=0}^{+\infty}\frac{T^{2\,n}}{C^2}\,{}_1 F_1(n+1,1,-|\alpha|^2)=e^{\frac{1}{2}\,|\alpha|^2}\,e^{-\,|\alpha|^2}\sum_{n=0}^{+\infty}\frac{T^{2\,n}}{C^2}\,{}_1 F_1(-n,1,|\alpha|^2)\nonumber\\
=&e^{-\frac{1}{2}\,|\alpha|^2}\,\sum_{n=0}^{+\infty}\frac{T^{2\,n}}{C^2}\,L_n(|\alpha|^2)\nonumber\\
=&e^{\frac{1}{2}\,|\alpha|^2}\,\frac{1}{C^2\,(1-T^2)}e^{-\frac{T^2\,|\alpha|^2}{1-T^2}},
\end{align}
which gives us the final, simple and compact result
\begin{align}\label{third:useful:expression:appendix}
J_\alpha=&e^{-\frac{1}{2}\,\cosh(2\,r)\,|\alpha|^2}.
\end{align}
In the above computations we have introduced the Confluent hypergeometric function ${}_1F_1(a,b;z)$ and the Laguerre polynomial $L_n(z)$. We have used the fundamental property ${}_1 F_1(a,b;z)=\exp[z]\,{}_1 F_1(b-a,b;-z)$ and the relation $L_n(z)={}_1 F_1(-n,1;z)$. We have used the \textit{generating function} for the Laguerre polynomial to be able to go to the last line of \eqref{calculation:one:ok} from the second-to-last one. 

We are also interested in $\tilde{J}_\alpha:=\sum_{n=0}^{+\infty}T^{2n}/C^2\langle n|\exp[\alpha\,\hat{a}^\dag-\alpha^*\,\hat{a}]\,\hat{a}\,|n\rangle$. We can proceed as above and find
\begin{align}\label{calculation:two:ok}
\tilde{J}_\alpha=&\sum_{n=0}^{+\infty}\frac{T^{2\,n}}{C^2}\,\langle n|\exp[\alpha\,\hat{a}^\dag-\alpha^*\,\hat{a}]\,\hat{a}\,|n\rangle\nonumber\\
=&e^{\frac{1}{2}\,|\alpha|^2}\sum_{n=0}^{+\infty}\sqrt{n}\,\frac{T^{2\,n}}{C^2}\langle n|e^{-\alpha^*\,\hat{a}}\,e^{\alpha\,\hat{a}^\dag}\,|n-1\rangle\nonumber\\
=&e^{\frac{1}{2}\,|\alpha|^2}\sum_{n,p,q=0}^{+\infty}\sqrt{n}\,\frac{T^{2\,n}}{C^2}\frac{(-\alpha^*)^p\,\alpha^q}{p!\,q!}\langle n|\hat{a}^p\,\hat{a}^{\dag q}\,|n-1\rangle\nonumber\\
=&e^{\frac{1}{2}\,|\alpha|^2}\sum_{n,p,q=0}^{+\infty}\sqrt{n}\,\frac{T^{2\,n}}{C^2}\frac{(-\alpha^*)^p\,\alpha^q}{p!\,q!}\,\sqrt{\frac{(p+n)!}{n!}}\,\sqrt{\frac{(q+n-1)!}{(n-1)!}}\,\langle n+p|n+q-1\rangle\nonumber\\
=&\alpha\,e^{\frac{1}{2}\,|\alpha|^2}\sum_{\overset{p=0}{n=1}}^{+\infty}\sqrt{n}\,\frac{T^{2n}}{C^2}\frac{(-|\alpha|^2)^p}{p!\,p!}\,\frac{(p+n)!}{\sqrt{n!\,(n-1)!}}=\alpha\,e^{\frac{1}{2}\,|\alpha|^2}\sum_{n=1}^{+\infty}\frac{n\,T^{2\,n}}{C^2}\,{}_1 F_1(n+1,2,-|\alpha|^2)\nonumber\\
=&\alpha\,e^{\frac{1}{2}\,|\alpha|^2}\sum_{m=0}^{+\infty}\frac{(m+1)\,T^{2(m+1)}}{C^2}\,{}_1 F_1(m+2,2,-|\alpha|^2)\nonumber\\
=&\alpha\,e^{\frac{1}{2}\,|\alpha|^2}\,T^2\,e^{-\,|\alpha|^2}\sum_{m=0}^{+\infty}\frac{T^{2\,m}}{C^2}\,{}_1 F_1(-m,2,|\alpha|^2)\nonumber\\
=&\alpha\,e^{-\frac{1}{2}\,|\alpha|^2}\,T^2\sum_{m=0}^{+\infty}\frac{T^{2\,m}}{C^2}\,L^{(1)}_m(|\alpha|^2)\nonumber\\
=&\alpha\,e^{-\frac{1}{2}\,|\alpha|^2}\,\frac{T^2}{C^2}\frac{1}{(1-T^2)^2}\,e^{-\frac{T^2}{1-T^2}\,|\alpha|^2}
\end{align}
which gives us the final, simple and compact result
\begin{align}\label{fourth:useful:expression:appendix}
\tilde{J}_\alpha=&\alpha\,\sinh^2r\,e^{-\frac{1}{2}\,\cosh(2\,r)\,|\alpha|^2}.
\end{align}
Note that here we have introduced the generalised Laguerre polynomials $L^{(q)}_a(z)$. Here as well, we have used the \textit{generating function} for the generalised Laguerre polynomial to go to the last line of \eqref{calculation:two:ok} from the second-to-last one. 

Finally, we want to compute the following $\tilde{L}_\alpha:=\sum_{l,l'=0}^{+\infty} T^{2(l+l')}/C^4\,\left|\langle l |\exp[\alpha\,\hat{a}^\dag-\alpha^*\,\hat{a}]|l'\rangle\right|^2$. We once more proceed using similar techniques as above, skip some passages and find
\begin{align}
\tilde{L}_\alpha=&\sum_{l,l'=0}^{+\infty} \frac{T^{2(l+l')}}{C^4}\,\left|\langle l |\exp[\alpha\,\hat{a}^\dag-\alpha^*\,\hat{a}]|l'\rangle\right|^2\nonumber\\
=&2\,e^{|\alpha|^2}\,\sum_{\overset{l=0}{q=1}}^{+\infty} \frac{T^{4\,l+2\,q}}{C^4}\,\frac{(l+q)!}{q!\,q!\,l!}
|\alpha|^{2\,q}\,\left|{}_1 F_1\left(l+q+1,q+1;-|\alpha|^2\right)\right|^2\nonumber\\
&+ e^{|\alpha|^2}\,\sum_{l=0}^{+\infty} \frac{T^{4\,l}}{C^4}\,
\left|{}_1 F_1\left(l+1,1;-\left|\alpha\right|^2\right)\right|^2\nonumber\\
=&2\,e^{-|\alpha|^2}\,\sum_{\overset{l=0}{q=1}}^{+\infty} \frac{T^{4\,l+2\,q}}{C^4}\,\frac{(l+q)!}{q!\,q!\,l!}
|\alpha|^{2\,q}\,\left|{}_1 F_1\left(-l,q+1;|\alpha|^2\right)\right|^2\nonumber\\
&+e^{-|\alpha|^2}\, \sum_{l=0}^{+\infty} \frac{T^{4\,l}}{C^4}\,
\left|{}_1 F_1\left(-l,1;\left|\alpha\right|^2\right)\right|^2\nonumber\\
=&2\,e^{-|\alpha|^2}\,\sum_{\overset{l=0}{q=1}}^{+\infty} \frac{t^{4\,l+2\,q}}{C^4}\,\frac{l!}{(l+q)!}\,\left|\alpha\right|^{2\,q}\,\left|L^{(q)}_{l}\left(\left|\alpha\right|^2\right)\right|^2+e^{-|\alpha|^2}\, \sum_{l=0}^{+\infty} \frac{T^{4\,l}}{C^4}\,\left|L_{l}\left(\left|\alpha\right|^2\right)\right|^2,
\end{align}
where ${}_1 F_1(a,b;x)$ is the Confluent hypergeometric function that satisfies ${}_1 F_1(a,b;x)=\exp[x]\,{}_1 F_1(b-a,b;-x)$, and the functions $L_a(x)$ and $L^{(q)}_a(x)$ are the Laguerre and generalised Laguerre polynomials respectively.

We now use the Hardy-Hille formula for Laguerre polynomials and obtain
\begin{align}
\tilde{L}_\alpha=&\frac{\exp\left[-\frac{1+T^4}{1-T^4}\left|\alpha\right|^2\right]}{\cosh (2\,r)}\,\left[2\,
\sum_{q=1}^{+\infty} I_q\left(2\,\frac{T^2}{1-T^4}\,\left|\alpha\right|^2\right)+I_0\left(2\,\frac{T^2}{1-T^4}\,\left|\alpha\right|^2\right)\right].
\end{align}
Finally, we use the fundamental Jacobi-Anger expansion, which can be recast as the identity $\exp[z\,\cos\theta]=I_0(z)+2\,\sum_{n=1}^\infty I_n(z)\,\cos(n\,\theta)$ for modified Bessel functions, and note that in our case we have $z=2\frac{T^2}{1-T^4}\,\left|\alpha\right|^2$ and $\theta=0$. Therefore
\begin{align}\label{fifth:useful:expression:appendix}
\tilde{L}_\alpha=&\frac{\exp\left[-\frac{1}{\cosh (2\, r)}\left|\alpha\right|^2\right]}{\cosh (2\,r)}.
\end{align}

%----------------------------------------------------------------------------------------------------------------------------------------------------------------------------------------------------------------------------%
\section{Reduced state of the resonators}\label{reduced:resonator:state:section}
%----------------------------------------------------------------------------------------------------------------------------------------------------------------------------------------------------------------------------%
We want to compute the final reduced state $\hat{\rho}_{\textrm{m}}$ of the mechanical resonators. The reduced state is defined by $\hat{\rho}_{\textrm{m}}:=\textrm{Tr}_{\textrm{Phot}}(\hat{\rho}_{\textrm{NL}}(t))$.
We assume that the initial state is $\hat{\rho}_0=\hat{\rho}_{\mathrm{c}}(0)\otimes\hat{\rho}_{\mathrm{m}}(0)$. We have
\begin{align}\label{nonlinear:redced:mechanical:state:appendix}
\hat{\rho}_{\textrm{m}}(t)=&\sum_{n_1,...,n_N}\langle n_1,...,n_N|\,\hat{U}_\textrm{NL}\,\hat{\rho}_0\,\hat{U}_\textrm{NL}^\dag| n_1,...,n_N\rangle\nonumber\\
=&\sum_{n_1,...,n_N}\langle n_1,...,n_N|\,\hat{\rho}_{\mathrm{c}}(0)| n_1,...,n_N\rangle\,\hat{D}_{\{n_k\}}\,\hat{\rho}_{\mathrm{m}}(0)\,\hat{D}_{\{n_k\}}^\dag\nonumber\\
=&\sum_{\overset{\{n_k\}}{k\in\mathcal{I}}} p_{\{n_k\}}\,\hat{D}_{\{n_k\}}\,\hat{\rho}_{\textrm{m}}(0)\,\hat{D}_{\{n_k\}}^\dag,
\end{align}
where we have introduced $\sum_{\overset{\{n_k\}}{k\in\mathcal{I}}}:=\sum_{n_1,n_2,...,n_N}$ for $N$ modes that belong to the set of all possible combinations of excitations $\mathcal{I}$, while $\sum_{k\in\mathcal{I}}\,P_k:=P_1+P_2+...+P_N$ for any $k$-dependent quantities $P_k$. We have also introduced
\begin{align}
| n_1,...,n_N\rangle:=&| n_1\rangle\otimes...\otimes| n_N\rangle\nonumber\\
p_{\{n_k\}}:=&\langle n_1,...,n_N|\hat{\rho}_{\mathrm{c}}(0)| n_1,...,n_N\rangle\nonumber\\
\hat{D}_{\{n_k\}}:=&\prod_p\,e^{-i\sum_p\,F^{(p)}_\textrm{m}\,\hat{b}^\dag_p\hat{b}_p}\,e^{-i\sum_p\left(F^{(p)}_++\sum_{k\in\mathcal{I}}n_k\,F^{(p,+)}_k\right)\,\hat{B}^{(p,+)}}\,e^{-i\sum_p\left(F^{(p)}_-+\sum_{k\in\mathcal{I}}n_k\,F^{(p,-)}_k\right)\,\hat{B}^{(p,-)}}.\nonumber\\
=&\prod_p e^{i\theta_p}\,\exp\left[-i\sum_p\,F^{(p)}_\textrm{m}\,\hat{b}^\dag_p\hat{b}_p\right]\,\exp\left[\left(F^{(p)*}+\sum_{k\in\mathcal{I}}n_k\,F^{(p)*}_k\right)\,\hat{b}^\dag_p-\textrm{h.c.}\right].
\end{align}
Finally, we have also introduced $F^{(p)}:=F^{(p)}_++i\,F^{(p)}_-$ and $F^{(p)}_k:=F^{(p,+)}_k+i\,F^{(p,-)}_k$. The phases $e^{i\theta_p}$ are irrelevant since they cancel out in the expression \eqref{nonlinear:redced:mechanical:state}.
Note that we have $\textrm{Tr}(\hat{\rho}_{\textrm{m}}(t))=\sum_{\overset{\{n_k\}}{k\in\mathcal{I}}} p_{\{n_k\}}=1$ as expected.

%----------------------------------------------------------------------------------------------------------------------------------------------------------------------------------------------------------------------------%
\section{Computing mixedness for initial coherent/thermal states of field modes/resonators}\label{summing:mixedness:section}
%----------------------------------------------------------------------------------------------------------------------------------------------------------------------------------------------------------------------------%
We want to compute an expression for $\text{Tr}_{\{m_{k}\},\{n_k\}}$ which is defined as
\begin{align}\label{applied:expression:linear:entropy:intermediate}
\text{Tr}_{\{m_{k}\},\{n_k\}}:=&\textrm{Tr}\left(\hat{D}_{\{m_{k}\}}^\dag\,\hat{D}_{\{n_k\}}\,\rho_{\textrm{m}}(0)\,\hat{D}_{\{n_k\}}^\dag\,\hat{D}_{\{m_{k}\}}\,\rho_{\textrm{m}}(0)\right),
\end{align}
where
\begin{align}
\hat{D}_{\{m_{k}\}}^\dag\,\hat{D}_{\{n_k\}}=&\prod_p e^{i\theta_p'}\,\exp\left[-i\sum_p\,F^{(p)}_\textrm{m}\,\hat{b}^\dag_p\hat{b}_p\right]\,\exp\left[i\,\Delta^{(p)}_{\{n_k,m_{k}\}}\,\hat{b}^\dag_p-\textrm{h.c.}\right]
\end{align}
and $\Delta^{(p)}_{\{n_k,m_{k}\}}:=\sum_{k\in\mathcal{I}}\left(n_k-m_k\right)F^{(p)}_k$ for simplicity of presentation. Again, the phases $e^{i\theta_p'}$ are irrelevant.

Our goal can be easily reached using the useful result \eqref{fifth:useful:expression:appendix}. We first note that we have to consider many $\tilde{L}_{\Delta^{(p)}_{\{n_k,m_{k}\}}}$ and $r_p$. We then note the important expression
\begin{align}
\text{Tr}_{\{m_{k}\},\{n_k\}}:=&\prod_p\,\tilde{L}_{i\,\Delta^{(p)}_{\{n_k,m_{k}\}}}.
\end{align}
This easily gives us
\begin{align}\label{applied:expression:linear:entropy:intermediate:final:appendix}
\text{Tr}_{\{m_{k}\},\{n_k\}}=&\prod_p \frac{\exp\left[-\frac{1}{\cosh (2\, r_p)}\left|\Delta^{(p)}_{\{n_k,m_{k}\}}\right|^2\right]}{\cosh (2\,r_p)}.
\end{align}

%----------------------------------------------------------------------------------------------------------------------------------------------------------------------------------------------------------------------------%
\subsection{Computing mixedness for initial coherent/thermal states of field modes/resonators}\label{summing:mixedness:special:case:one:mode:many:resonators:subsection}
%----------------------------------------------------------------------------------------------------------------------------------------------------------------------------------------------------------------------------%
Here we need to compute the mixedness \eqref{applied:full:expression:linear:entropy:main:result:final}. This requires us to compute the simpler contribution
\begin{align}
\Lambda_\alpha=e^{-2\,|\mu|^2}\,\sum^{+\infty}_{nm}\frac{|\mu|^{2\,(n+m)}}{n!\,m!}\,e^{-\alpha\,(n-m)^2}.
\end{align}
The calculations follow here
\begin{align}
\Lambda_\alpha=&e^{-2\,|\mu|^2}\,\sum^{+\infty}_{nm}\frac{|\mu|^{2\,(n+m)}}{n!\,m!}\,e^{-\alpha\,(n-m)^2}\nonumber\\
=&e^{-2\,|\mu|^2}\,\left[2\,\sum^{+\infty}_{n>m}\frac{|\mu|^{2\,(n+m)}}{n!\,m!}\,e^{-\alpha\,(n-m)^2}+\sum_n\,\frac{|\mu|^{4\,n}}{n!\,n!}\right]\nonumber\\
=&e^{-2\,|\mu|^2}\,\left[2\,\sum^{+\infty}_{\overset{n=0}{d=1}}\frac{|\mu|^{4\,n+2\,d}}{n!\,(n+d)!}\,e^{-\alpha\,d^2}+\sum_n\,\frac{|\mu|^{4\,n}}{n!\,n!}\right]\nonumber\\
=&e^{-2\,|\mu|^2}\,\left[2\,\sum^{+\infty}_{d=1}\,I_d(2\,|\mu|^2)\,e^{-\alpha\,d^2}+I_0(2\,|\mu|^2)\right].
\end{align}
Using the Jacobi-Anger expansion for the modified Bessel functions we obtain
\begin{align}
\Lambda_\alpha=&1-2\,e^{-2\,|\mu|^2}\,\sum^{+\infty}_{d=1}\,I_d(2\,|\mu|^2)\,\left(1-e^{-\alpha\,d^2}\right).
\end{align}

%----------------------------------------------------------------------------------------------------------------------------------------------------------------------------------------------------------------------------%
\section{Time evolution in the periodically driven linearised regime}\label{linearised:driven:section}
%----------------------------------------------------------------------------------------------------------------------------------------------------------------------------------------------------------------------------%
Here we compute the time evolution induced by the linearized Hamiltonian $\hat {H}_{\textrm{lin}}$ when only one cavity mode $\tilde{k}$ is present. It reads
\begin{align}\label{main:linearised:time:dependent:Hamiltonian:to:decouple:appendix}
\hat {H}_{\textrm{lin}}=& \hbar\,\omega_{\textrm{c},\tilde{k}} \delta\hat  a^\dagger_{\tilde{k}} \delta\hat a_{\tilde{k}} + \sum_p\hbar\,\omega_{\textrm{m},p} \,\hat b_p^\dagger \hat b_p+\sum_p\hbar\,g_{\tilde{k} p}\,\alpha_{\tilde{k}}^2\,\hat{B}^{(+)}_p+\sum_{p}\hbar\,\alpha_{\tilde{k}}\,g_{\tilde{k}p}\,\hat{A}^{(+)}_{\tilde{k}}\,\hat{B}^{(+)}_p.
\end{align}
This Hamiltonian induced the time evolution through the operator
\begin{align}\label{time:evolution:operator:linearised:regime:appendix}
\hat{U}_{\textrm{lin}}(t)=&\hat{U}_0(t)\,\hat{U}_1(t)\,\hat{U}_\mathrm{D}(t),
\end{align}
where we have defined 
\begin{align}\label{time:evolution:operators:list:linearised:regime:appendix}
\hat{U}_0(t)=&\exp\left[-i\,\left( \omega_{\textrm{c},\tilde{k}} \delta\hat   a_{\tilde{k}}^\dagger \delta\hat a_{\tilde{k}} + \sum_p\omega_{\textrm{m},p} \,\hat b_p^\dagger \hat b_p\right)\,t\right]\nonumber\\
\hat{U}_1(t)=&\hat{U}_0(t)\,\overset{\leftarrow}{\mathcal{T}}\,\exp\left[-i/\hbar\,\int_0^t\,dt'\,\hat{H}_1\right].
\end{align}
The Hamiltonians in \eqref{time:evolution:operators:list:linearised:regime:appendix} read
\begin{align}\label{main:linearised:time:dependent:Hamiltonian:to:decouple:convenient:appendix}
\hat{H}_1(t)=&+\sum_{p}\hbar\,\alpha_{\tilde{k}}\,g_{\tilde{k}p}\,\cos((\omega_{\textrm{c},\tilde{k}}+\omega_{\textrm{m},p})\,t)\,\left(\delta\hat{a}^\dag_{\tilde{k}}\hat{b}^\dag_p+\delta\hat{a}_{\tilde{k}}\hat{b}_p\right)+\sum_{p}\hbar\,\alpha_{\tilde{k}}\,g_{\tilde{k}p}\,\sin((\omega_{\textrm{c},\tilde{k}}+\omega_{\textrm{m},p})\,t)\,i\,\left(\delta\hat{a}^\dag_{\tilde{k}}\hat{b}^\dag_p-\delta\hat{a}_{\tilde{k}}\hat{b}_p\right)\nonumber\\
&+\sum_{p}\hbar\,\alpha_{\tilde{k}}\,g_{\tilde{k}p}\,\cos((\omega_{\textrm{c},\tilde{k}}-\omega_{\textrm{m},p})\,t)\,\left(\delta\hat{a}^\dag_{\tilde{k}}\hat{b}_p+\delta\hat{a}_{\tilde{k}}\hat{b}^\dag_p\right)+\sum_{p}\hbar\,\alpha_{\tilde{k}}\,g_{\tilde{k}p}\,\sin((\omega_{\textrm{c},\tilde{k}}-\omega_{\textrm{m},p})\,t)\,i\,\left(\delta\hat{a}^\dag_{\tilde{k}}\hat{b}_p-\delta\hat{a}_{\tilde{k}}\hat{b}^\dag_p\right)\nonumber\\
&+\sum_p\hbar\,g_{\tilde{k}p}\,\alpha_{\tilde{k}}^2\,\cos(\omega_{\textrm{m},p}\,t)\,\hat{B}^{(+)}_p+\sum_p\hbar\,g_{\tilde{k}p}\,\alpha_{\tilde{k}}^2\,\sin(\omega_{\textrm{m},p}\,t)\,\hat{B}^{(-)}_p
\end{align}
Here we will assume that the couplings $g_{\tilde{k}p}(t)$ are modulated periodically. In this work, the choice of modulation is not restricted by any physical constraint, therefore we can choose $g_{\tilde{k}p}(t)=g_{\tilde{k}p}\,(1+\kappa\,\sin(\omega_\mathrm{d}\,t))$ or $g_{\tilde{k}p}(t)=g_{\tilde{k}p}\,(1+\kappa\,\cos(\omega_\mathrm{d}\,t))$ as simple modulations. It is easy to see that the choice of modulation will ``select'' a particular term in the expression for $\hat{H}_1(t)$. What we mean by this is that
\begin{align}
\sin(\omega_\mathrm{d}\,t)\,\sin((\omega_{\textrm{c},\tilde{k}}\pm\omega_{\textrm{m},p})\,t)=& \frac{1}{2}\left[\cos((\omega_\mathrm{d}-(\omega_{\textrm{c},\tilde{k}}\pm\omega_{\textrm{m},p}))\,t)-\cos((\omega_\mathrm{d}+(\omega_{\textrm{c},\tilde{k}}\pm\omega_{\textrm{m},p}))\,t)\right]\nonumber\\
\cos(\omega_\mathrm{d}\,t)\,\cos((\omega_{\textrm{c},\tilde{k}}\pm\omega_{\textrm{m},p})\,t)=& \frac{1}{2}\left[\cos((\omega_\mathrm{d}+(\omega_{\textrm{c},\tilde{k}}\pm\omega_{\textrm{m},p}))\,t)+\cos((\omega_\mathrm{d}-(\omega_{\textrm{c},\tilde{k}}\pm\omega_{\textrm{m},p}))\,t)\right],
\end{align}
which have particular, or \textit{resonant}, behavior when $\omega_\mathrm{d}=\omega_{\textrm{c},\tilde{k}}\pm\omega_{\textrm{m},p}$. Assuming that all frequencies are positive and that $\omega_{\textrm{c},\tilde{k}}>\omega_{\textrm{m},p}$, we then have 
\begin{align}
\sin(\omega_\mathrm{d}\,t)\,\sin((\omega_{\textrm{c},\tilde{k}}\pm\omega_{\textrm{m},p})\,t)=& \frac{1}{2}\left[1-\cos(2\,\omega_\mathrm{d}\,t)\right]\nonumber\\
\cos(\omega_\mathrm{d}\,t)\,\cos((\omega_{\textrm{c},\tilde{k}}\pm\omega_{\textrm{m},p})\,t)=& \frac{1}{2}\left[1+\cos(2\,\omega_\mathrm{d}\,t)\right]
\end{align}
when $\omega_\mathrm{d}=\omega_{\textrm{c},\tilde{k}}\pm\omega_{\textrm{m},p}$.

In the following we will consider both cases separately.

%----------------------------------------------------------------------------------------------------------------------------------------------------------------------------------------------------------------------------%
\subsection{Time evolution in the periodically driven linearised regime: mode-mixing drive}
%----------------------------------------------------------------------------------------------------------------------------------------------------------------------------------------------------------------------------%
Here we consider the case where $\omega_\mathrm{d}=\omega_{\textrm{c},\tilde{k}}-\omega_{\textrm{m},\tilde{p}}$ for a specific oscillator $\tilde{p}$ and a drive of the form $g_{\tilde{k}p}(t)=g_{\tilde{k}p}\,(1+\kappa\,\sin(\omega_\mathrm{d}\,t))$. This means that, after long enough time, the effective Hamiltonian contributions will be coming from the term
\begin{align}\label{main:linearised:time:dependent:Hamiltonian:to:decouple:convenient:squeezingappendix}
\hat{H}_1(t)\sim&\frac{1}{2}\,\hbar\,\alpha_{\tilde{k}}\,\kappa\,g_{\tilde{k}\tilde{p}}\,i\,\left(\delta\hat{a}^\dag_{\tilde{k}}\hat{b}_{\tilde{p}}-\delta\hat{a}_{\tilde{k}}\hat{b}_{\tilde{p}}^\dag\right).
\end{align}
This implies that $\hat{U}_1(t)\sim\exp\bigr[\alpha_{\tilde{k}}\,\kappa\,g_{\tilde{k}\tilde{p}}\,\bigl(\delta\hat{a}^\dag_{\tilde{k}}\hat{b}_{\tilde{p}}-\delta\hat{a}_{\tilde{k}}\hat{b}^\dag_{\tilde{p}}\bigr)\bigl]$ after enough time.

This also means that it is easy to compute the time evolution through the linearised Hamiltonian in this regime, which after enough time reads
\begin{align}\label{main:linearised:time:dependent:Hamiltonian:to:decouple:convenient:modemixing:appendix}
\hat{U}_{\textrm{lin}}(t)\,\delta\hat{a}_{\tilde{k}}\,\hat{U}_{\textrm{lin}}(t)=&e^{-i\,\omega_{\textrm{c},\tilde{k}}\,t}\left[\delta\hat{a}_{\tilde{k}}\,\cos(\alpha_{\tilde{k}}\,\kappa\,g_{\tilde{k}\tilde{p}}\,t)+\hat{b}_{\tilde{p}}\,\sin(\alpha_{\tilde{k}}\,\kappa\,g_{\tilde{k}\tilde{p}}\,t)\right]\nonumber\\
\hat{U}_{\textrm{lin}}^\dag(t)\,\hat{b}_{\tilde{p}}\,\hat{U}_{\textrm{lin}}(t)=&e^{-i\,\omega_{\textrm{m},\tilde{p}}\,t}\left[\hat{b}_{\tilde{p}}\,\cos(\alpha_{\tilde{k}}\,\kappa\,g_{\tilde{k}\tilde{p}}\,t)-\delta\hat{a}_{\tilde{k}}\,\sin(\alpha_{\tilde{k}}\,\kappa\,g_{\tilde{k}\tilde{p}}\,t)\right].
\end{align}
Notice that, as expected for a resonant drive scenario, the mode-mixing angle $\theta_{\textrm{m},\tilde{p}}$ between the cavity mode and resonator, reads $\theta_{\textrm{m},\tilde{p}}=\alpha_{\tilde{k}}\,\kappa\,g_{\tilde{k}\tilde{p}}\,t$ and increases with time. 

%----------------------------------------------------------------------------------------------------------------------------------------------------------------------------------------------------------------------------%
\subsection{Time evolution in the periodically driven linearised regime: squeezing drive}
%----------------------------------------------------------------------------------------------------------------------------------------------------------------------------------------------------------------------------% 
Here we consider the case where $\omega_\mathrm{d}=\omega_{\textrm{c},\tilde{k}}+\omega_{\textrm{m},\tilde{p}}$ for a specific oscillator $\tilde{p}$ and a drive of the form $g_{\tilde{k}p}(t)=g_{\tilde{k}p}\,(1+\kappa\,\sin(\omega_\mathrm{d}\,t))$. This means that, after long enough time, the effective Hamiltonian contributions will be coming from the term
\begin{align}\label{main:linearised:time:dependent:Hamiltonian:to:decouple:convenient:squeezing:appendix}
\hat{H}_1(t)\sim&\frac{1}{2}\,\hbar\,\alpha_{\tilde{k}}\,\kappa\,g_{\tilde{k}\tilde{p}}\,i\,\left(\delta\hat{a}^\dag_{\tilde{k}}\hat{b}^\dag_{\tilde{p}}-\delta\hat{a}_{\tilde{k}}\hat{b}_{\tilde{p}}\right).
\end{align}
This implies that $\hat{U}_1(t)\sim\exp\bigl[\alpha_{\tilde{k}}\,\kappa\,g_{\tilde{k}\tilde{p}}\,\bigl(\delta\hat{a}^\dag_{\tilde{k}}\hat{b}^\dag_{\tilde{p}}-\delta\hat{a}_{\tilde{k}}\hat{b}_{\tilde{p}}\bigr)\bigr]$ after enough time.

This also means that it is easy to compute the time evolution through the linearised Hamiltonian in this regime, which after enough time reads
\begin{align}\label{main:linearised:time:dependent:Hamiltonian:to:decouple:convenient:squeezingappendix}
\hat{U}_{\textrm{lin}}(t)\,\delta\hat{a}_{\tilde{k}}\,\hat{U}_{\textrm{lin}}(t)=&e^{-i\,\omega_{\textrm{c},\tilde{k}}\,t}\left[\delta\hat{a}_{\tilde{k}}\,\cosh(\alpha_{\tilde{k}}\,\kappa\,g_{\tilde{k}\tilde{p}}\,t)+\hat{b}_{\tilde{p}}^\dag\,\sinh(\alpha_{\tilde{k}}\,\kappa\,g_{\tilde{k}\tilde{p}}\,t)\right]\nonumber\\
\hat{U}_{\textrm{lin}}^\dag(t)\,\hat{b}_{\tilde{p}}\,\hat{U}_{\textrm{lin}}(t)=&e^{-i\,\omega_{\textrm{m},\tilde{p}}\,t}\left[\hat{b}_{\tilde{p}}\,\cosh(\alpha_{\tilde{k}}\,\kappa\,g_{\tilde{k}\tilde{p}}\,t)+\delta\hat{a}_{\tilde{k}}^\dag\,\sinh(\alpha_{\tilde{k}}\,\kappa\,g_{\tilde{k}\tilde{p}}\,t)\right].
\end{align}
Notice that, as expected for a resonant drive scenario, the squeezing $r_{\textrm{m},\tilde{p}}$ between the cavity mode and resonator, reads $r_{\textrm{m},\tilde{p}}=\alpha_{\tilde{k}}\,\kappa\,g_{\tilde{k}\tilde{p}}\,t$ and increases with time. 

\end{document}